\documentclass[
reprint,
superscriptaddress,
amsmath,amssymb,
aps,
prb,
longbibliography
]{revtex4-1}

\usepackage{graphicx,epsf}
\usepackage{float}
\usepackage{xcolor}
\usepackage{bm}

\begin{document}

\title{Scattering of topological kink-antikink states in bilayer graphene
structures
}

\author{Nassima Benchtaber}
\affiliation{Institute of Interdisciplinary Physics and Complex Systems IFISC 
(CSIC-UIB), E-07122 Palma, Spain} 
\author{David S\'anchez}
\affiliation{Institute of Interdisciplinary Physics and Complex Systems IFISC 
(CSIC-UIB), E-07122 Palma, Spain} 
\affiliation{Department of Physics, University of the Balearic Islands, 
E-07122 Palma, Spain}
\author{Lloren\c{c} Serra}
\affiliation{Institute of Interdisciplinary Physics and Complex Systems IFISC 
(CSIC-UIB), E-07122 Palma, Spain} 
\affiliation{Department of Physics, University of the Balearic Islands, 
E-07122 Palma, Spain}

\begin{abstract}
Gapped bilayer graphene can support the presence of intragap states
due to kink gate potentials applied to the graphene layers.
Electrons in these states display valley-momentum locking,
which makes them attractive for 
topological valleytronics. Here, we show that kink-antikink
local potentials enable modulated scattering of topological currents.
We find that the kink-antikink coupling leads to anomalous steps in the junction
conductance. Further, when the constriction detaches from the propagating
modes, forming a loop, the conductance reveals the system energy spectrum.
Remarkably, these kink-antikink devices can also work as valley filters
with tiny magnetic fields
by tuning a central gate.
\end{abstract}

\maketitle

\section{Introduction}

For many years, there has been considerable interest in providing with reliable platforms that can create, manipulate and detect qubits. Eventually, these quantum information processing tasks are to be supplemented with protected communication channels to transmit quantum states between distant sites. Graphene has emerged as an excellent candidate in scalable solid-state architectures due to its ultra long decoherence times for spin qubits and its ability to host additional isospin (valley) degrees of freedom \cite{Trau07,Pereira07,Ryc07,Recher10,Gun11}. 
These emerge as the $K$ and $K'$ points from the Dirac cones in the reciprocal space of the graphene hexagonal lattice.

However, pristine graphene lacks a bandgap, which handicaps potential applications of this material for nanoelectronics. This circumstance can be surpassed with the employment of two graphene sheets (hereafter, bilayer graphene or BLG)~\cite{Mcan13,rozhkov16}. Interlayer coupling in a Bernal stacking structure (common also to graphite) generates a huge band splitting of the order of 380 meV, although two bands still remain degenerate at the neutrality point. Further application of a perpendicular electric field creating a potential difference 
between the two layers finally lifts the electronic degeneracy~\cite{Zhang09,Overweg18}.
The gap thus opened can now be used to design tunnel barriers
and quantum point contacts~\cite{Over18,Kraf18}.
Unlike monolayer point contacts
that are fabricated by etching~\cite{Ter16,Cle19} and show trapped states
due to edge roughness, BLG quantum wires display clear conductance quantization steps. Further, if two of these barriers are connected in
a series the device works as a quantum dot~\cite{Eich18,Kurzmann19,Banszerus20,Banszerus21}. 
Therefore, robust spin or valley qubits can form in BLG dots showing a discrete spectrum.

An even more exciting possibility arises in BLG systems. When the 
perpendicular electric field becomes inhomogeneous by changing its sign in different 
regions of the BLG, the domain wall separating the two opposite fields holds topological states propagating next to the wall (edge states)~\cite{Mar08,Zarenia11}. 
These types of domain wall and propagating states are known as kink and kink states, respectively.
The topological character originates from the field induced band inversion and confinement. Interestingly, the valley index remains a good quantum number~\cite{Zhang13} and as a consequence the kink states become chiral with different valleys traveling in opposite directions along the kink~\cite{Lon15}. 
This is a consequence of the chiral symmetry that relates states with
opposite energies, valleys and propagation directions.
Crucially, such propagating states exist at zero magnetic field,
in which case time-reversal symmetry is fulfilled.
Hence, if electrons are injected from the side into a straight kink using a small dc bias, the output flux becomes valley polarized. 
If the structure is built zero dimensional like a dot, the bound states are valley degenerate but chiral.
These topological states can even show Luttinger behavior~\cite{Kil10} or become massless Dirac modes~\cite{Kil11} in the presence of interactions or periodic potentials, respectively. 

\begin{figure}[t]
\begin{center} 
\includegraphics[width=0.45 \textwidth, trim = 1.5cm 3.5cm 1.5cm 
2.5cm,clip]{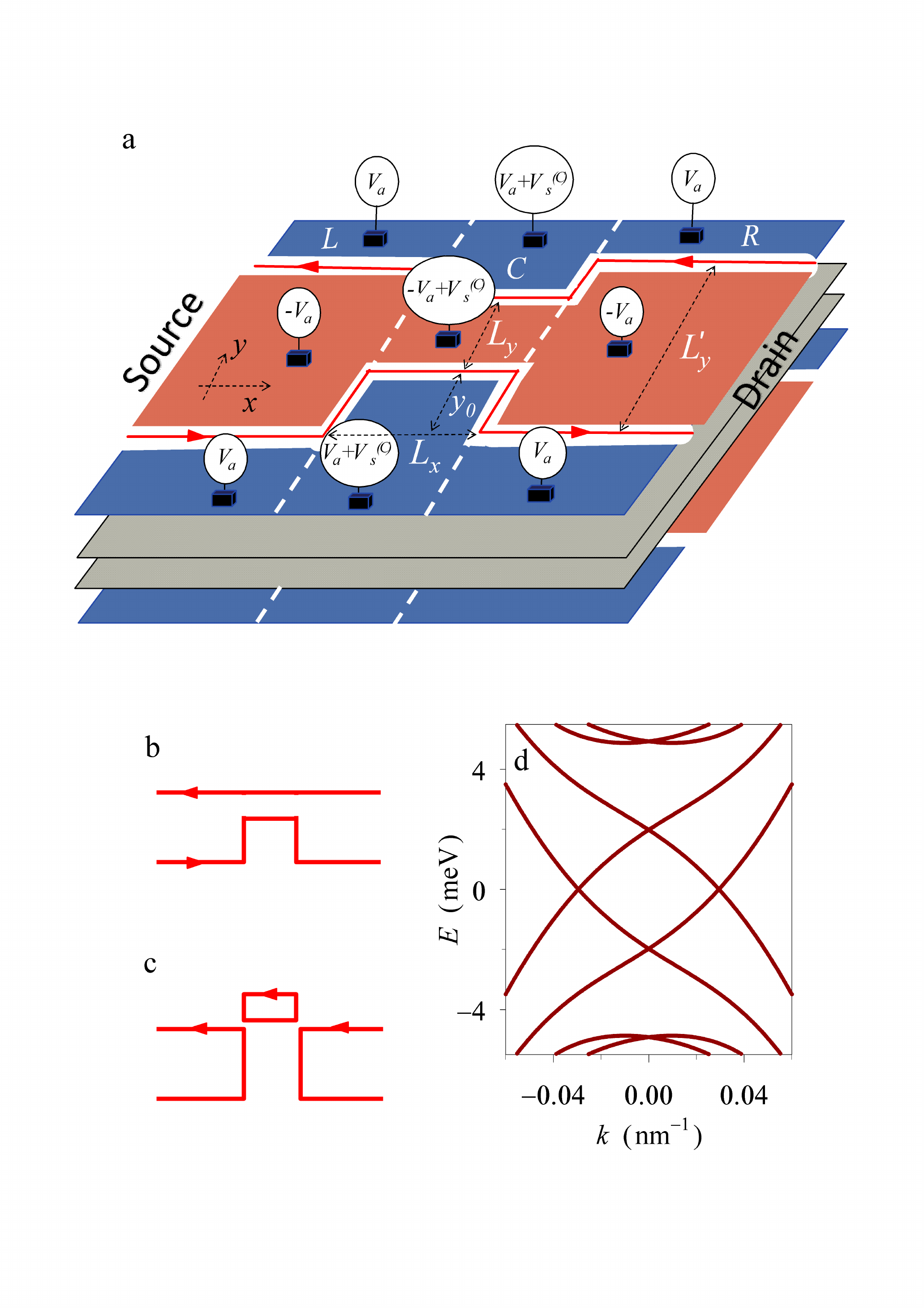}
\end{center}
\caption{
(a)
Schematic of a bilayer graphene kink-antikink system with lead regions ($L$ and $R$)
and a central scatterer ($C$).
The electric field direction on the  
graphene layers (gray sheets) is controlled  
by the voltages applied to the nine top and bottom gates (blue and orange regions).
The values of the $V_a$ applied potentials to the lower gates (not shown) are reversed with respect to the top gates.
The field inversion between the blue and orange regions creates an interface
(white region) where topological modes emerge.
Additionally, a voltage $V_s^{(C)}$ is applied only to the central region.
Red lines represent these propagating topological modes for the valley $K$.
$K'$ modes are obtained by reversing the arrows. Current is generated from
the left (source) terminal to the right (drain) terminal.
Dimensions and position of the central scatterer
are given by  $L_x$, $L_y$ and $y_0$, while the asymptotic separation of the two kinks is $L'_y$.
(b,c) Selected configurations representing a 
particular constriction (b) and 
side loop (c), the latter formed when $y_0>L'_y$.
(d)~Energy bands of a translationally invariant kink-antikink wire of width $L'_y=300\,{\rm nm}$,
kink potential height $V_a=10\,{\rm meV}$ and potential smoothness $s=38\,{\rm nm}$.
Each band is fourfold degenerate (spin and valley) in the absence of magnetic fields.
}
\label{Fsk}
\end{figure}

The challenge then is how to probe and manipulate these unique kink states. This can be achieved with pairs of gates
whose voltage is tuned independently in both the top and bottom layers~\cite{Li16,Chen20}. Experimentally, a conductance of $4e^2/h$ is observed,~\cite{Lon15,Li16,Chen20} 
which demonstrates the presence of two current-carrying spin-degenerate valley-polarized modes. Alternate combinations of dual gates
can be implemented for guiding these modes in valley valves and beam splitters~\cite{Xia07,Qiao11,Li18}.
However, signatures of disorder are detected, inducing backscattering and intervalley mixing. What is needed is a controlled source of backscattering that would allow to shape ballistic beam splitters for, e.g., topological valleytronic interferometry~\cite{Cheng18}. Here, we show that this is possible with a careful distribution of top/bottom gate pairs, enabling the formation of a lateral constriction over {\em two} 
parallel kinks, as sketched in Fig.\ \ref{Fsk}a. 
The two kinks have symmetric changes in the field polarity and opposite 
propagation directions for a given valley,
thus one kink being the antikink of the other \cite{Mar08,Zarenia11}. 
We below demonstrate that a kink-antikink constriction is able to modulate the 
transmission electrostatically, what paves the 
way for the fabrication of topological quantum point contacts and wires.

Importantly, the setup can be slightly rearranged to also furnish bound states when the kink-antikink constriction (Fig.\ \ref{Fsk}b)
transforms into a side loop (Fig.\ \ref{Fsk}c).
Previous works on similar BLG islands assumed a sharp-potential kink~\cite{xavier10}
or infinite-mass boundary conditions~\cite{Dacosta14} that create rings exhibiting Aharonov-Bohm energy levels. However, these are closed systems
and as such their properties would be difficult to examine in an experiment. In contrast, our loops are weakly coupled to the external (side) kinks,
topological as well. We show that the measured conductance peak 
pattern is caused by the system level distribution.
Therefore, the setup is most suitable 
for doing spectroscopy of chiral bound states.

Let us discuss in more detail our proposal, as illustrated in Fig.~\ref{Fsk}a,
and highlight our main findings. The system consists of a BLG with the same gate distribution
in both the upper and lower graphene layers (gray sheets).
The applied potentials to the lower gates, not shown in Fig.~\ref{Fsk}a, reverse the
values for $V_a$ with respect to the top gates but keep the same 
value for $V_s^{(C)}$, as detailed below in Sec.\ \ref{model}.
The changes in electric field orientation
occur in the white interfaces
defining, respectively, the topological kink and antikink that
form the quasi-one dimensional (1D) propagating channels. 
The arrows in Fig.\ \ref{Fsk}a qualitatively indicate electron propagation for a given valley on the BLG planes
when the kink and antikink are well separated.
$x$ is the transport direction, the edge states are confined along $y$ and
the direction perpendicular to the graphene layers is denoted with $z$ (not shown here).
Electronic motion is determined by chirality  
due to valley-momentum locking (we only depict states from valley $K$). 

A narrow constriction in the central region allows for a controlled
transmission of the injected beams. 
In our parametrization (Fig.\ \ref{Fsk}a), 
a constriction corresponds to having $L_y<L'_y$ and $y_0\leq L'_y-L_y$.
A particular example for
a constriction 
is shown in Fig.\ \ref{Fsk}b  for $y_0=L'_y-L_y$, 
although other configurations are possible.
We find that for narrow constrictions the precise value of 
$y_0$ is not relevant and the conductance shows \textit{anomalous} 
steps as the central potential is varied.
If $y_0 \geq L'_y$ or $y_0\leq -L_y$
the constriction becomes a loop that detaches from the left and right channels; 
Fig.\ \ref{Fsk}c shows a case for $y_0 > L'_y$. 
For narrow loops the conductance
displays \textit{resonant} peaks as a function of central potential, 
their location giving information about the energy levels inside the 
loop.
Altogether, the structure is a remarkable playground for 
electrical transport
studies of both propagating and localized topological valley states.

\section{Model}
\label{model}

We use an effective eight-component
model, valid for low energies near the Dirac points of the BLG 
crystalline band structure. The Hamiltonian reads~\cite{Mcan13,rozhkov16}
\begin{eqnarray}
 H &=& v_F \left(p_x- \hbar  \frac{y}{l_z^2}\right) \tau_z \sigma_x
 + v_F\, p_y \sigma_y \nonumber\\
 &+& \frac{t}{2}\, \left(\,\lambda_x \sigma_x +\lambda_y\sigma_y\,\right) 
 +V_s + V_a\, \lambda_z\; ,
\label{eq1}
 \end{eqnarray}
with three characteristic pseudospins 
(valley $\tau_{xyz}$, sublattice $\sigma_{xyz}$ and 
layer $\lambda_{xyz}$)
described by corresponding Pauli matrices
while $p_x$ and $p_y$ are momentum operators.
Two of the model paremeters are intrinsic of BLG, namely,
the graphene Fermi velocity $\hbar v_F=660\,{\rm meV}\,{\rm nm}$
and the interlayer coupling $t=380\,{\rm meV}$. 
Then, 
$l_z=\sqrt{\hbar /eB}$ is the magnetic length for an external magnetic field $B$
described in the Landau gauge
whereas $V_s$ and $V_a$ are respectively the symmetric and asymmetric potentials
applied to the layers. 
For uniform potentials, $V_s$ is just a global energy shift
while $V_a$ is a displacement energy that opens a gap in the BLG spectrum.
The inhomogenous system of Fig.\ \ref{Fsk}a has
position dependent potentials  
$V_a(y)$ and $V_s(x)$, with transitions between plateau values 
defined by the gates.

The time reversal $\Theta$ and chiral-symmetry  ${\cal C}$ operators 
and transformations relevant to our system read
\begin{equation}
\begin{array}{llll}
\Theta = i\tau_y {\cal K}\; &\Rightarrow  & \Theta^2=-1\; , & \Theta H(B)\Theta =-H(-B)\;, \\
{\cal C} = \sigma_x\tau_x\lambda_y\;
&\Rightarrow
&
{\cal C}^2=1\; ,
&
{\cal C}H{\cal C} = -H
\; ,
\end{array}
\end{equation}
where ${\cal K}$ refers to complex conjugation.
The symmetry transformations on a state 
$|Ekv\rangle$ with a given energy, momentum and valley
are given by $\Theta|Ekv\rangle \propto |E\bar{k}\bar{v}\rangle$ and
${\cal C}|Ekv\rangle \propto |\bar{E}k\bar{v}\rangle$. 

We first discuss the spectrum that arises from Eq.~\eqref{eq1} 
for a translationally invariant kink-antinkink system
at $B=0$ and $V_s=0$.
The absence of a central region in Fig.\ \ref{Fsk}a 
can be represented by 
$L_x=0$ or, alternatively, by $y_0=0$
and $L'_y=L_y$.
In this case, states propagate along $x$ and are characterized by 
a real wave number $k$, i.e., $p_x\to \hbar k$ in Eq.\ (\ref{eq1}).
Whenever $V_a=V_a (y)$ changes its sign the gap
is inverted and as a consequence four topological states per valley appear at 
each kink~\cite{Mar08,Zarenia11}.
These correspond to the branches seen around zero energy in Fig.~\ref{Fsk}d.
The states above $E=4.2$~meV are extended states that do not remain attached to the kinks
in contrast to the topological states.
Further, the energy bands in Fig.~\ref{Fsk}d
are not bounded either from below or from above since Eq.~\eqref{eq1} describes Dirac fermions.
We also note that $H$ is both valley diagonal (so that each valley can be independently treated in a four-component subspace)
and diagonal in the real spin basis. However, whereas all states are hereafter degenerate for spins up and down,
the spectrum is not valley degenerate but obeys $E(k,\tau_z\!\to\! 1)=E(-k,\tau_z\!\to\!-1)$
due to time reversal symmetry.
As a consequence, kink-antikink currents are valley unpolarized.
Later, we will remark that a magnetic field breaks time reversal symmetry and thus
valley polarizations can be observed in the measured conductance.

In our calculations, the kink potentials
vary smoothly in $y$ by means of a diffusivity $s$ 
(see App.~\ref{sec_pot}
for details of the potential modeling). 
This smoothness becomes important 
when the kink-antikink separation is small,
i.e., the constriction in Fig.\ \ref{Fsk}b or the loop in Fig.\ \ref{Fsk}c. Then, $s$
couples the kink states running on the two sides, a mechanism that is
eventually responsible for the transmission modulation.
Along the transport direction $x$,
the potential interfaces are considered sharp.
This assumption is well justified since the mode wavelength $\lambda$
is much larger than the characteristic length $l_a$ for inversion of the static potentials. Electrostatic modeling in bilayer graphene\cite{Li16,Chen20} yields an estimate $l_a < 50$ nm, while in our calculations we
typically have $\lambda \gtrsim 300$ nm.

We next consider the inhomogenous situation with $L$, $C$ and $R$ 
regions along the transport direction $x$, sketched in Fig.\ \ref{Fsk}a.
The distribution of applied potentials is seen 
in Fig.\ \ref{Fsk}a for the top layer. Gates on the bottom layer 
have reversed $V_a$ and the same central shift $V_s^{(C)}$.
We solve the scattering problem in the presence of either the constriction or the loop using complex 
band structure methods \cite{Serra13,Osca19}.
A survey of this method
is given in App.\ \ref{CBS}.
The technique is especially well suited to describe piecewise homogenous potentials in topological systems.
For each region $a=L,C,R$ in Fig.~\ref{Fsk}a, a large set of complex wavenumbers and eigenstates 
$\{k^{(a)},\phi_k^{(a)}\}$ is determined by exact diagonalization \cite{arpack}. 
These sets of solutions 
are then properly matched at the interfaces between central ($C$) and side regions ($L,R$).
The ensuing linear system of equations  
yields the transmission amplitudes $t_{n'n}$ from input mode $n$ to output mode $n'$. The electric conductance is then determined 
by the two-terminal formula $G=(2e^2/h)\sum_{nn'}{|t_{n'n}|^2}$,
where spin degeneracy is already taken into account 
and we assume zero temperature
(the experiments in Ref.~\onlinecite{Over18} are done at a very low temperature of 1.7~K).
The set of complex wavenumbers and wavefunctions
of each region is obtained with a finite difference discretization of
a 1D equation depending only on $y$ since the $x$ dependence disappears thanks to
the homogeneity of each region along the transport direction.
This 1D character enables an accurate numerical resolution for large numbers of $y$ grid points, 
while no grid in $x$ is needed.

The use of grid 
discretization methods for Dirac-like problems leads to the infamous Fermion doubling problem~\cite{Susskind77,Nielsen81,Lewe12},
which introduces spurious replica states. These 
are characterized by very short wavelength oscillations, strongly fluctuating
from one grid point to the next. Similar replicas 
are obtained in our approach when calculating the complex band structure of each region $\{k^{(a)},\phi_k^{(a)}\}$. We filter out the replicas by 
coarse graining, performing an average with the right or left neighboring
point and neglecting those states whose norm is affected by coarse graining.
For dense grids, we easily arrive at an unambiguous identification 
of the physical states, which need to be smooth on the grid by definition. 
Thus, the Fermion doubling problem does not affect the linear system that determines
the conductance since $G$ is based only on the sets of 
previously filtered solutions  $\{k^{(a)},\phi_k^{(a)}\}$ and no further spatial 
grid is required near the interfaces.

\section{Results}

We study two systems formed with the gate distribution and geometry depicted in Fig.~\ref{Fsk}a:
(i) when 
$0<y_0\le L_y'-L_y$
propagating modes
can exist within the central area and a quantum point contact behavior is expected (Fig.~\ref{Fsk}b);
(ii) when $y_0>L_y'$ a loop detaches from the left and right leads (Fig.~\ref{Fsk}c)
and we will consequently find quantum resonance
effects. In both devices,
a key parameter is the symmetric 
potential $V_s^{(C)}$ in the central region, which acts as an effective local probe
allowing energy spectroscopy of the constriction.

\begin{figure}[t]
\begin{center}
\includegraphics[width=0.5\textwidth,trim=1.cm 10.75cm 1.cm 2.2cm,clip]{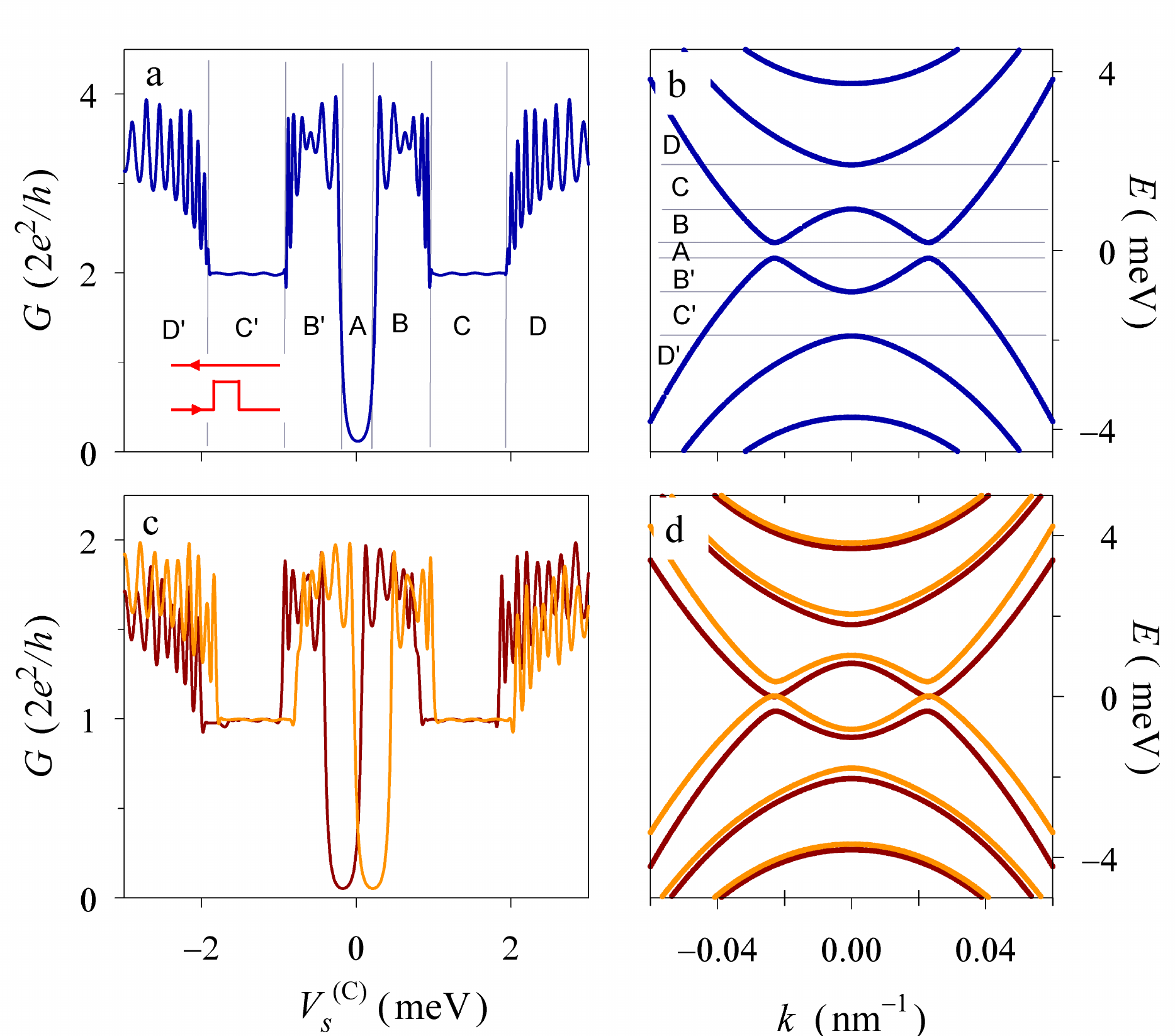}
\end{center}
\caption{
Results for a narrow constriction (inset in panel a) with
$L_y=100\,{\rm nm}$, $y_0=200\,{\rm nm}$ and a
Fermi energy $E=0.02\, {\rm meV}$.
(a) Conductance for $L_x=1\,{\mu}{\rm m}$ as a function of the 
central potential $V_s^{(C)}$.
(b) Energy bands for a kink-antikink wire having 
the same parameters of the constriction.
The capital letters and horizontal lines indicate the correspondence with the conductance ranges of panel a.
(c,d) Same as panels a and b, respectively, with a 
magnetic field of $B= 50\,{\rm mT}$. The two colors mark the two different valleys.
}
\label{Fig2}
\end{figure}

\subsection{Quantum point contacts}

We first present results for 
a constriction corresponding to a
narrow point contact with a kink-antikink separation of $100\,{\rm nm}$.
This value is compatible with the width of presently available BLG point contacts~\cite{Over18,Kraf18}.
We set the kink diffusivity to $s = 38\,{\rm nm}$,
which is taken from the electric potential distribution
in dual split gate BLG devices~\cite{Li16,Chen20},
and assume an almost vanishing Fermi energy $E=0.02\,{\rm meV}$,
close to the charge neutrality point. 
The results are not strongly affected by changes around this value, as 
long as higher energy modes in the asymptotic kink-antikink 
are not activated and remain far from the Fermi energy (cf.\ Fig.\ \ref{Fsk}d).

The conductance for a $1\,\mu{\rm m}$-long 
constriction (hereafter the wire) as a function of the central potential 
is shown in Fig.~\ref{Fig2}a.
We observe that $G$ is strongly suppressed around $V_s^{(C)}=0$ (region A).
This is in principle a surprise since the presence of the edge states
at $E=0$ in Fig.~\ref{Fsk}d would imply a fully transparent constriction.
However, when we plot in Fig.~\ref{Fig2}b the wire band structure
we notice that the topological bands display an absolute gap (for any $k$)
in region A (details of this gap are discussed in App.~\ref{sec_gap}).
The kink potentials in the constriction couple
the edge states, leading to an almost complete backscattering
and hence a reduction of the conductance. 
$G$ does not reach zero because
the electrons can traverse the constriction by tunnel effect, which yields in any case a tiny value for $G$.
Then, as $V_s^{(C)}$ increases the energy exceeds the gap
and we find in region B two propagating states with positive velocity,
per valley and spin.
It follows that $G$ quickly reaches the quantized value of 
$8e^2/h$.
If $V_s^{(C)}$ is further enhanced we enter region C, where a
single mode is only allowed, thus bringing $G$ down to an anomalous step of 
$4e^2/h$.
Finally, larger values of $V_s^{(C)}$ approach us
into region D, where another mode starts to contribute and $G$
grows again. The oscillations seen in Fig.~\ref{Fig2}a are due to
quantum interference of several modes coexisting in the wire.
We can thus conclude that there is a remarkable correspondence
between $G$ and the wire energy bands. In fact, 
the particle-hole symmetry of Fig.~\ref{Fig2}b
implies that $G(V_s^{(C)})=G(-V_s^{(C)})$, as found in Fig.~\ref{Fig2}a.
We also point out that for nonzero
temperatures the conductance curves will be thermal smeared.

\begin{figure}
\begin{center}
\includegraphics[width=0.5\textwidth,trim=1.5cm 11cm 1cm 2.5cm,clip]{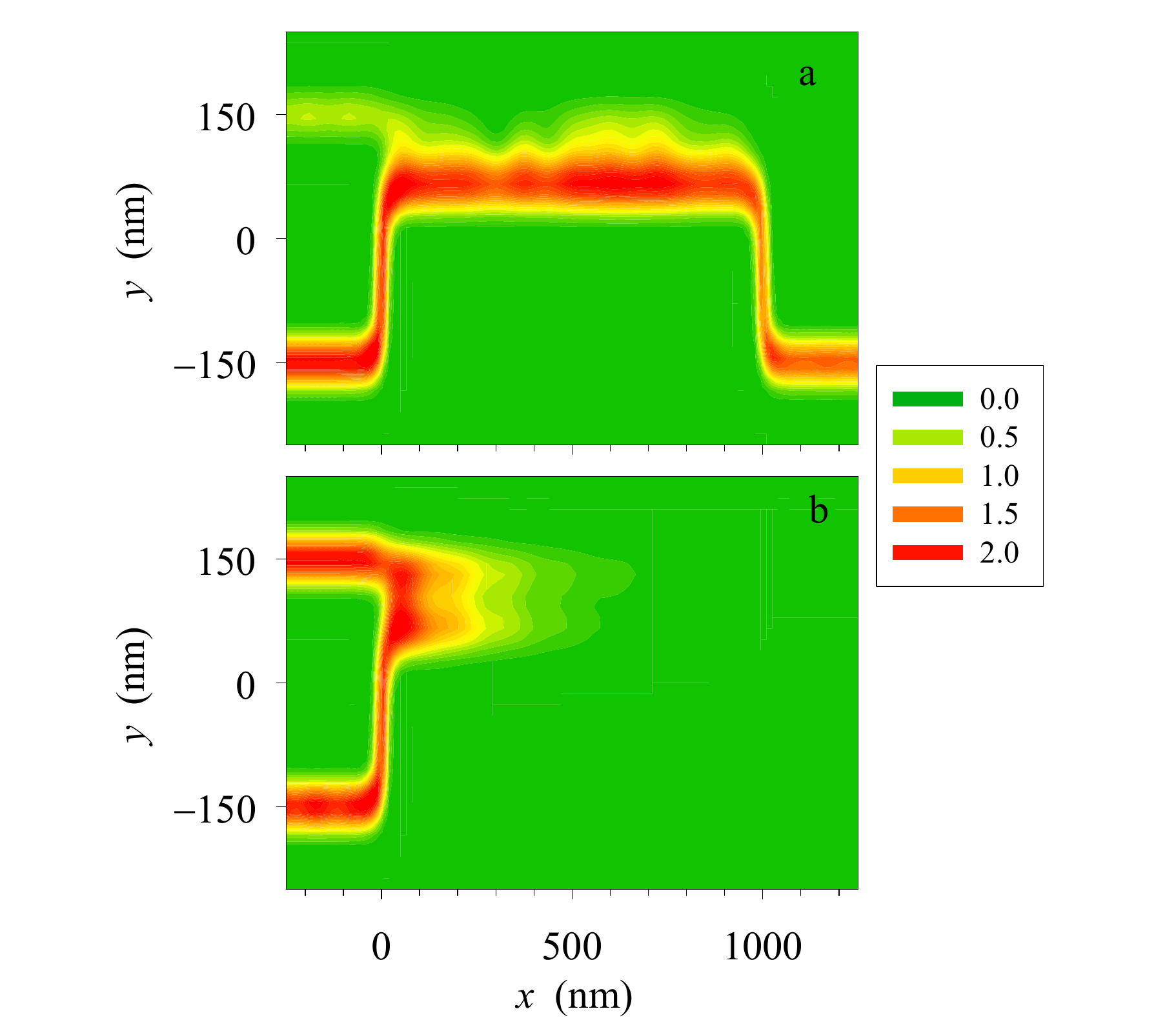}
\end{center}
\caption{%
Density distribution, in arbitrary units, 
for the scattering states
corresponding to the conductance curves of Fig.~\ref{Fig2}c.
Panels a and b are for the two different valleys $K$ and $K'$,
respectively. 
Parameters: $L_x=1000\;{\rm nm}$, 
$L_y=100\; {\rm nm}$,
$L'_y=300\; {\rm nm}$,
$y_0=200\; {\rm nm}$,
$B=50\;{\rm mT}$, $E=0.02\;{\rm meV}$ and $V_s^{(C)}=0.2\;{\rm meV}$.
}
\label{fig_7S}
\end{figure}

A small magnetic field, in the mT range, suffices to 
yield large effects on the constriction conductance 
even if the valley splitting of the energy bands is small (Fig.\ \ref{Fig2}c). 
Here, we choose to separately display each valley contribution to the conductance.
Accordingly, the scale of $G$ reduces a factor $2e^2/h$ as compared with Fig.\ \ref{Fig2}a.
We find that the conductance
shifts in opposite directions for the two valleys $\tau_z\to\pm 1$, making it 
possible the creation of highly polarized valley currents, where one valley component is 
essentially blocked while the other is transmitted.
The valley split bands are shown in Fig.\ \ref{Fig2}d.
This behavior can be also seen with a single kink due to valley-momentum locking.
However, if we wish to invert the current valley polarization with a kink we would need to revert the extended lateral gates defining the kink whereas
Fig.\ \ref{Fig2}c shows the interesting 
possibility of switching the valley polarization by simply changing $V_s^{(C)}$,
leaving both the lateral gates defining the kinks 
and the magnetic field fixed.
Therefore,
our system would work as an {\em electrically tunable, fully reversible valley filter} using tiny magnetic fields. 
Notice that BLG valley filters based on nontopological states 
require much larger fields, in the tesla range~\cite{Par19}.

This is better seen in Fig.~\ref{fig_7S}, where we plot the density distribution 
of the Fermi-energy scattering states
when electrons are injected from the source terminal (left side).
The two valleys (Fig.~\ref{fig_7S}a and Fig.~\ref{fig_7S}b)  contribute differently
since the magnetic field is finite. While for valley $K$ electrons impinge from the bottom left
kink (Fig.~\ref{fig_7S}a), the opposite valley $K'$ electrons (Fig.~\ref{fig_7S}b) enter from the top left
kink. The former (latter) are mostly transmitted (reflected), giving rise to a valley polarized current
in the drain terminal (right side).

\begin{figure}
\begin{center}
\includegraphics[width=0.5\textwidth,trim=3.5cm 17.5cm 1.2cm 2.5cm,clip]{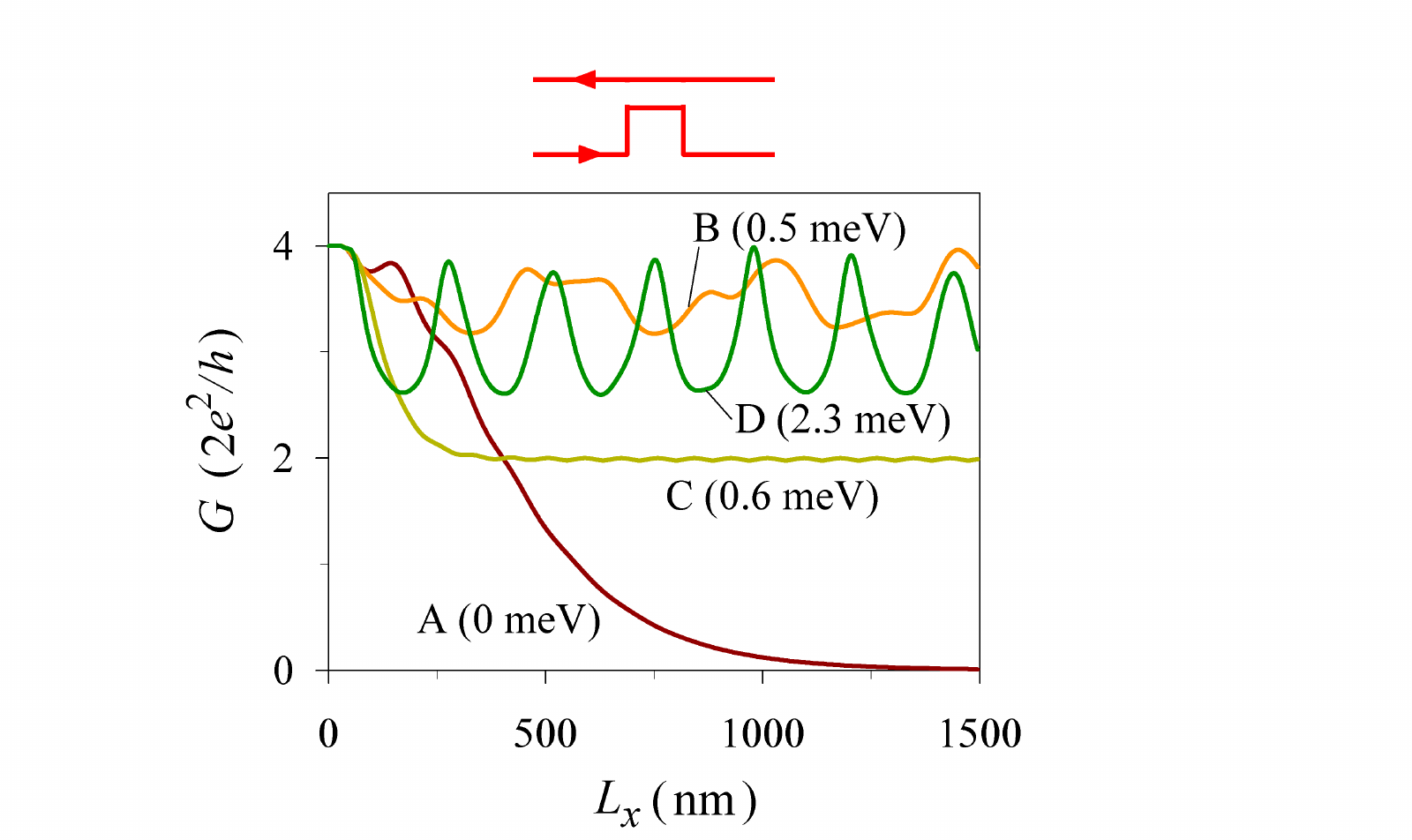}
\end{center}
\caption{
Dependence of the conductance on the constriction length $L_x$ for
the results of Fig.\ \ref{Fig2}a and
selected values of $V_s^{(C)}$, as given in parenthesis. 
A-D labels are used to indicate the same regions of 
Fig.~\ref{Fig2}a. 
} 
\label{fig_Lx}
\end{figure}

The dependence of $G$ on the constriction length $L_x$ (see Fig.\ \ref{fig_Lx})
further supports our interpretation. We display the conductance for $V_s^{(C)}$
corresponding to the four regions indicated in Fig.~\ref{Fig2}a.
In the gapped region A the conductance decays exponentially for large values 
of $L_x$, which agrees with a transport mechanism based on tunnel effect.
In region B the conductance shows an oscillatory
behavior up to arbitrarily large distances, implying a Fabry-Perot interference between propagating
modes in the central area.
The conductance becomes quantized at $4e^2/h$ in region C, which occurs when
the interfaces between the leads and the constriction becomes transparent.
Finally, in region D we recover the oscillatory behavior due to the activation of a new transport channel.
In all cases the role of quantum tunneling for small $L_x$ is clearly seen because the conductance
increases as $L_x$ shrinks to zero and scattering thus disappears.

\subsection{Side loops}

\begin{figure}
\begin{center}
\includegraphics[width=0.47\textwidth,trim=0.6cm 12.5cm 0.3cm 2.2cm,clip]{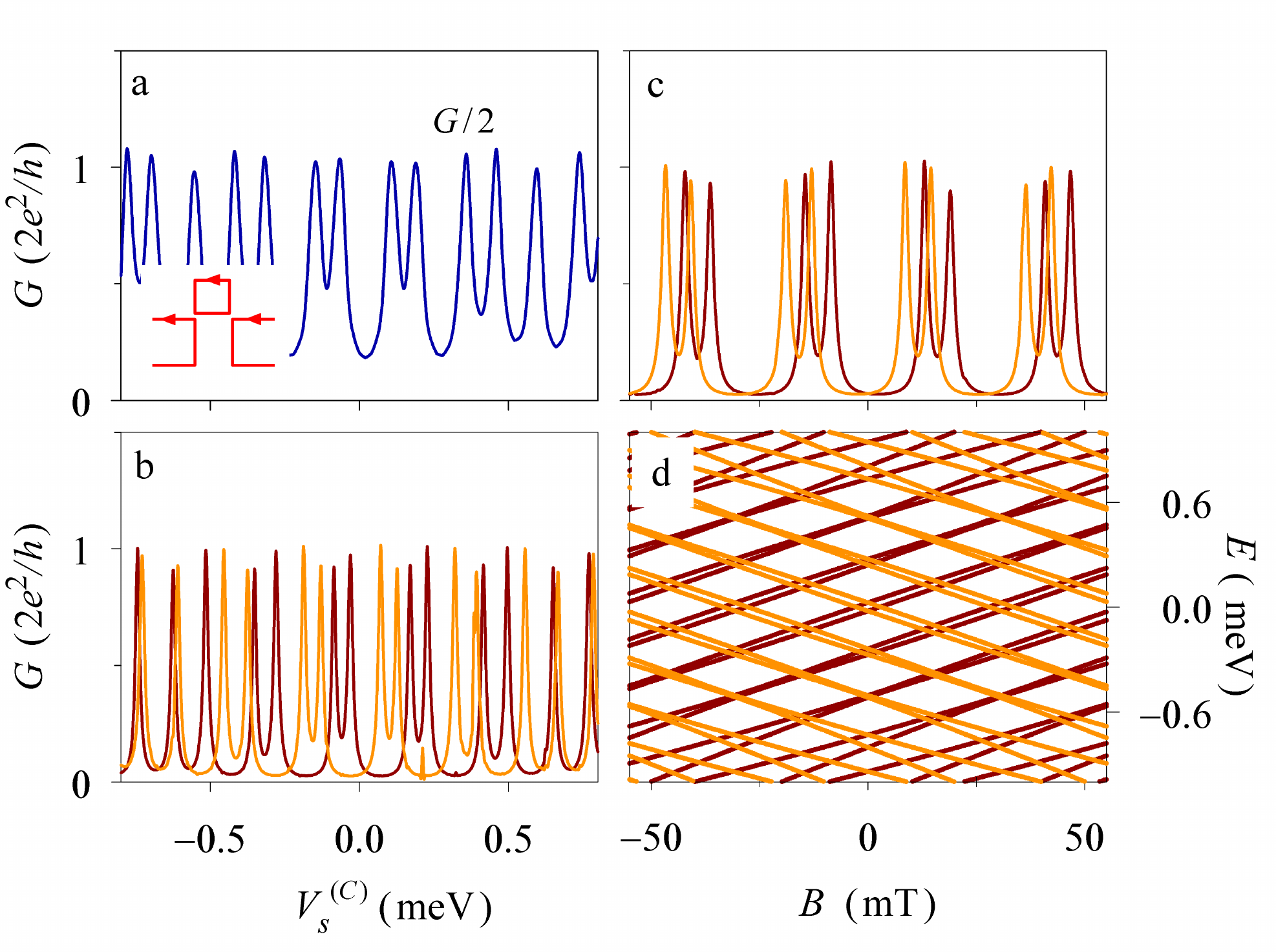}
\end{center}
\caption{Results for a side loop  
(inset in panel a) with $L_y=150\,{\rm nm}$, $y_0=310\,{\rm nm}$ and 
Fermi energy $E=0.02\; {\rm meV}$.
(a,b) Conductance for $L_x=1\,\mu{\rm m}$  as a function of the 
central potential $V_s^{(C)}$ for  $B=0$ ({\rm a}) and $B=50\,{\rm mT}$ (b).
(c) Conductance as a function of the magnetic field
for $V_s{(C)}=0$.
(d) Energy levels for the finite
loop $(L_x,L_y)= (1\,\mu{\rm m}, 150\,{\rm nm})$ 
as a function of the field. 
The two colors in panels b-d indicate the two different valleys.
The peak separation [around $0.04\,{\rm meV}$ in (b)]
could be resolved at low temperatures
$T\lesssim 0.5\, {\rm K}$.
}
\label{Fig3}
\end{figure}

Let us turn to the loops created as the gate position $y_0$ shown in 
Fig.\ {\ref{Fsk}a} increases.
Then, the edge states in the central region detach as illustrated in Fig.~\ref{Fsk}c.
Figure~\ref{Fig3}a shows in this case a conductance pattern that strongly differs from the wire system of Fig.~\ref{Fig2}a.
$G$ is characterized by resonant peaks that reach values of the order of $4e^2/h$
(we plot $G/2$ for convenience). Interestingly, these peaks
are causally correlated with the discrete levels in the closed loop. 
To see this, we plot in Fig.~\ref{Fig3}d the loop energy spectrum.
We find that the position of the conductance peaks agree, apart from a slight renormalization due to
the coupling with to external edge states, with the level positions. The particular peak structure
is highly sensitive to the loop dimensions $(L_x,L_y)$ due to quantum confinement. 

A small magnetic field splits the conductance peaks, as shown in Fig.\ \ref{Fig3}b
where we plot the valley resolved $G$ for $B=50$~mT.
It is noticeable that the peak
widths are significantly reduced in the presence of $B$, thus leading to 
smaller conductance minima; cf.\ Figs.\ \ref{Fig3}a and \ref{Fig3}c.
The conductance splitting is explained with the level behavior as a function of $B$ as shown in Fig.~\ref{Fig3}d. 
The field acts differently on the two valleys, thus raising (lowering) the energy for $\tau_z \to 1$ ($\tau_z \to -1$).
$B$-splitting of the two valleys is also present for the case of non topological bound states in graphene circular quantum dots,
discussed in Ref.\ \onlinecite{Recher09}, 
where states of the 
same angular momentum and opposite valleys show opposite dispersions 
at low fields. 
Besides the splitting, the spectrum in Fig.\ \ref{Fig3}d 
for topological loops  
shows a pattern of almost parallel lines for each valley,
reflecting a quantization condition of the topological states along the 
perimeter of the loop~\cite{xavier10}. 
Our results obey reciprocity, i.e., $G$ is unchanged when both $B$ and the valley index are simultaneously reversed, as can be seen 
in Fig.~\ref{Fig3}c.
It is also worth stressing that 
the valley and gate sensitivity allows, as in the constriction, switching
the valley polarization of the current by soley tuning the gate potential,
only that a finer tuning is needed in the detached loop in order
to hit the narrower peak maxima. 
Resolving the narrow peaks needs low temperatures to avoid thermal broadening, which we 
can estimate below $500\, {\rm mK}$ based on Fig.\ \ref{Fig3}.

\begin{figure}
\begin{center}
\includegraphics[width=0.5\textwidth,trim=1.5cm 7.cm 1cm 2.5cm,clip]{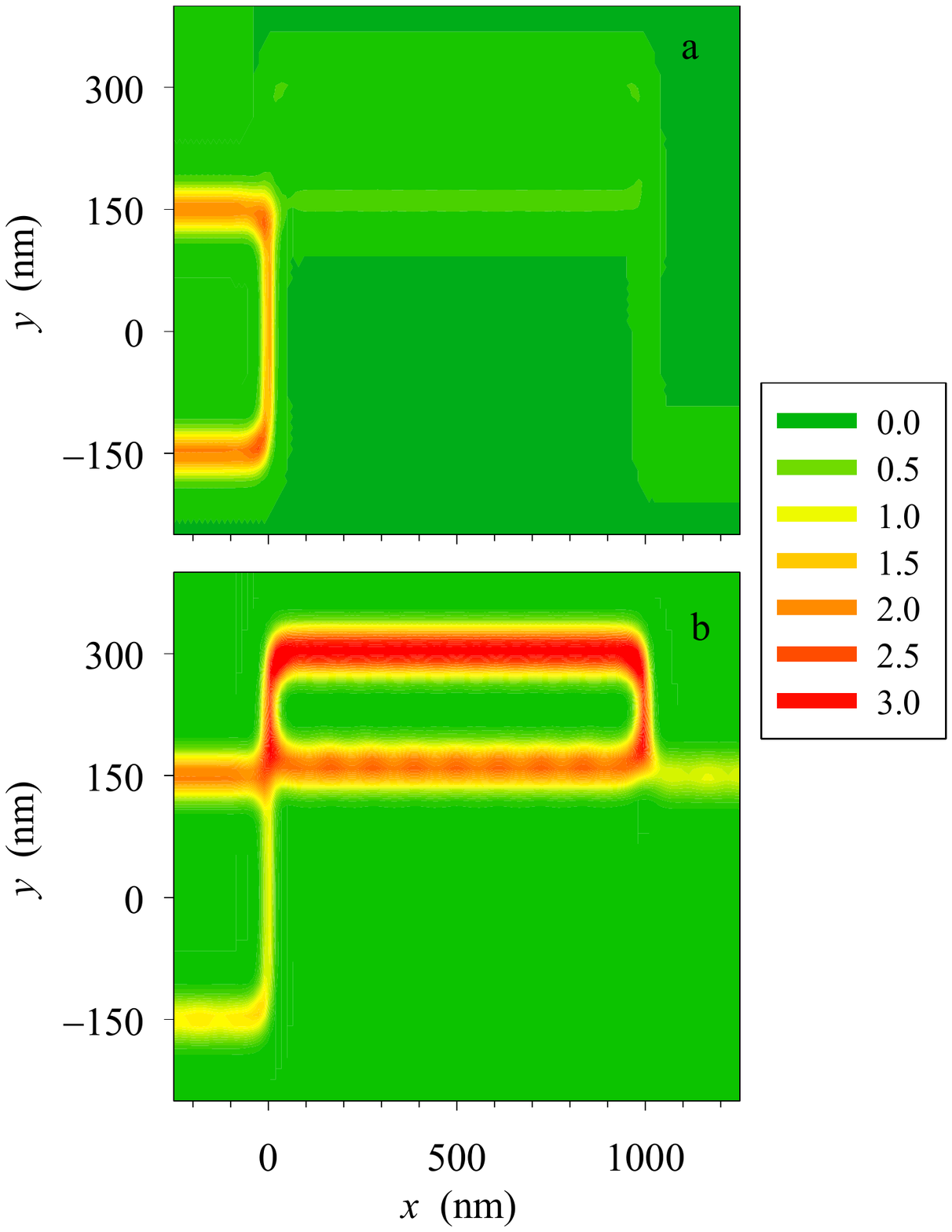}
\end{center}
\caption{%
Density distribution corresponding to the results of a side loop depicted in Fig.~\ref{Fig3}b.
The central gate potential is  
$V_s^{(C)}=0.3\,{\rm meV}$ and it  corresponds to 
a $K'$ valley  peak (light color) in Fig.~\ref{Fig3}b.
Panels a and b are for the two different valleys $K$ and $K'$,
respectively. 
Parameters: $L_x=1000\;{\rm nm}$,
$L_y=150\; {\rm nm}$,
$L'_y=300\; {\rm nm}$,
$y_0=310\; {\rm nm}$,
$B=50\;{\rm mT}$, $E=0.02\;{\rm meV}$ and $V_s^{(C)}=0.3\;{\rm meV}$.}
\label{fig_8S}
\end{figure}

Probability density distributions in space 
provide a more visual support for this valley switch effect (see Fig.~\ref{fig_8S}).
We note that just one valley is populating the loop (Fig.~\ref{fig_8S}b) while the other one is 
reflected (Fig.~\ref{fig_8S}a).
Thus, the valley-split resonant conductances of side loops in small magnetic fields imply high valley accumulations on the loop for specific gate potentials.

\begin{figure}
\begin{center}
\includegraphics[width=0.37\textwidth,trim=3.5cm 7.5cm 5cm 10.cm,clip]{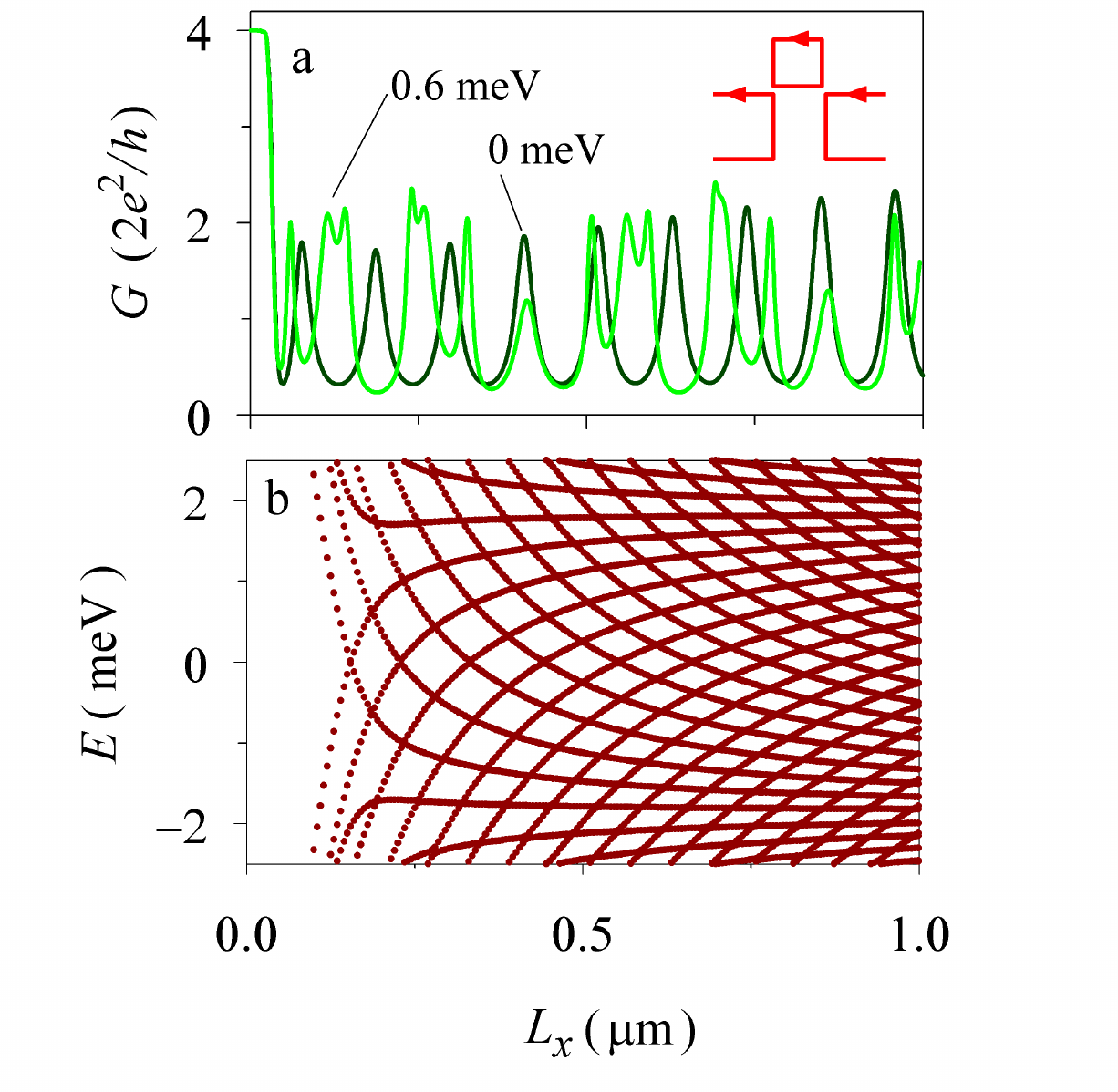}
\end{center}
\caption{%
(a) $L_x$ dependence of the conductance for two selected values of $V_s^{(C)}$ for the parameters of the 
$L_y=150\; {\rm  nm}$ 
side loop of Fig.\ \ref{Fig3}a.
(b) Energy levels of the  closed rectangular loop with $L_y=150\;{\rm nm}$ as a function of $L_x$.
}
\label{fig_3S}
\end{figure}

\begin{figure}
\begin{center}
\includegraphics[width=0.37\textwidth,trim=3.5cm 7.5cm 5cm 10cm,clip]{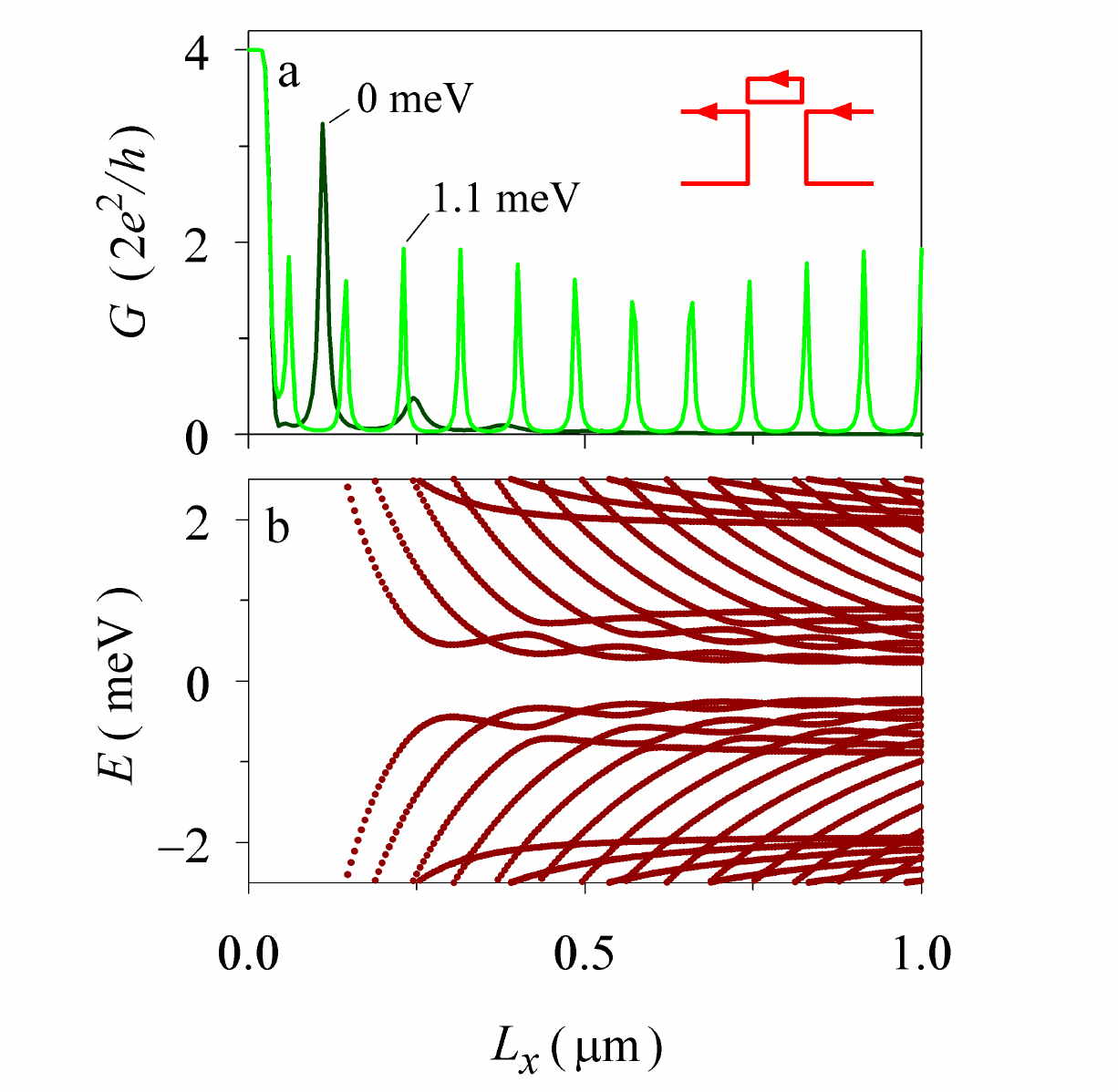}
\end{center}
\caption{%
(a) $L_x$ dependence of the conductance for a narrow loop ($L_y=100\;{\rm nm}$) and two
selected values of $V_s^{(C)}$ with the rest of the parameters as in Fig.\ \ref{fig_3S}.
(b) Energy levels of the closed loop with the same $L_y=100\;{\rm nm}$ as a function of $L_x$.
}
\label{fig_4S}
\end{figure}

We now briefly discuss the dependence on $L_x$. In Fig.~\ref{fig_3S} we show results for a wide loop
of $L_y=150\;{\rm nm}$ as in Fig.~\ref{Fig3}a. For 
$V_s^{(C)}$ corresponding to a
conductance valley in Fig.~\ref{Fig3}a the conductance curve in Fig.~\ref{fig_3S}a (black line) shows a single-mode regular spacing,
which is in agreement with the degeneracy due to level crossing for $E=0$ of Fig.~\ref{fig_3S}b.
In contrast, for  a $V_s^{(C)}$ value that generates a $G$ peak in Fig.~\ref{Fig3}a the conductance curve
shows multiple-mode spacings in  Fig.~\ref{fig_3S}a (light green line). 
Further, we get
accidental crossings at values of $L_x$ that lead to additional
factor-2 degeneracies in Fig.~\ref{fig_3S}b.
The case of a narrow
loop ($L_y=100\;{\rm nm}$) is shown in Fig.~\ref{fig_4S}. In contrast to the previous case,
the conductance valley curve is quenched as $L_x$ increases (black line in Fig.~\ref{fig_4S}a)
whereas the peak curve
exhibits a beating pattern overimposed to the peak sequence (green line).
Figure~\ref{fig_4S}b shows the energy levels when the narrow loop is closed. We observe regions with a bunching
of levels separated from others with regularly spaced levels. This is consistent with the
conductance peaks obtained in Fig.~\ref{fig_4S}a. Despite the fact that the conductance patterns
depend on the specific geometry of the loop (via $L_x$ and $L_y$), in all cases the conductance
peaks are correlated with the energy levels of the closed loop. Thus, the conductance serves
as an excellent tool to probe the internal structure of topologically bound states.

\section{Conclusions}

We have proposed a versatile nanodevice for topological studies
in quantum valley transport. Transmission manipulation is achieved
by means of a kink-antikink local potential that allows the formation
of (i) point contacts with anomalous quantized conductance and (ii)
side loops with chiral quasi-bound states.
The obtained conductance curves provide information
on the system energy spectrum.
For tiny magnetic fields we obtain a valley polarization effect
and this polarization is tunable with the gate potential.

Possible drawbacks of our proposal might be the effect of imperfections 
and misalignments of top and bottom gates in Fig.\ \ref{Fsk}a
as well as the presence of disorder due to impurities.
Our results, however, suggest 
robustness against small gate displacements  
since we take into account
smooth transition profiles of the asymmetric potential
with  diffusivity values as large as $s\approx 40\,{\rm nm}$. 
On the other hand, the chiral character of the low-energy 
valley-momentum-locked states 
offers protected transmission against backscattering by impurities that
conserve the valley degree of freedom~\cite{Li11}.

The illustrative examples considered in this work do not exhaust the capabilities
of the system, and more sophisticated setups could be envisaged.
Further, our model could be straightforwardly extended
to multivalley materials other than bilayer graphene such as silicene~\cite{Pan14},
sonic crystals~\cite{Lu17} and photonic platforms~\cite{Gao17}.

\acknowledgements
We acknowledge support from MINECO (Spain) Grant
No.\ MAT2017-82639, 
No.\ PID2020-117347GB-100,
MINECO/AEI/FEDER Mar\'{\i}a de
Maeztu Program for Units of Excellence MDM2017-0711.

\appendix

\section{Potential smoothness}\label{sec_pot}

\begin{figure}
\begin{center}
\includegraphics[width=0.5\textwidth,trim=3cm 13cm 1.5cm 8.2cm,clip]{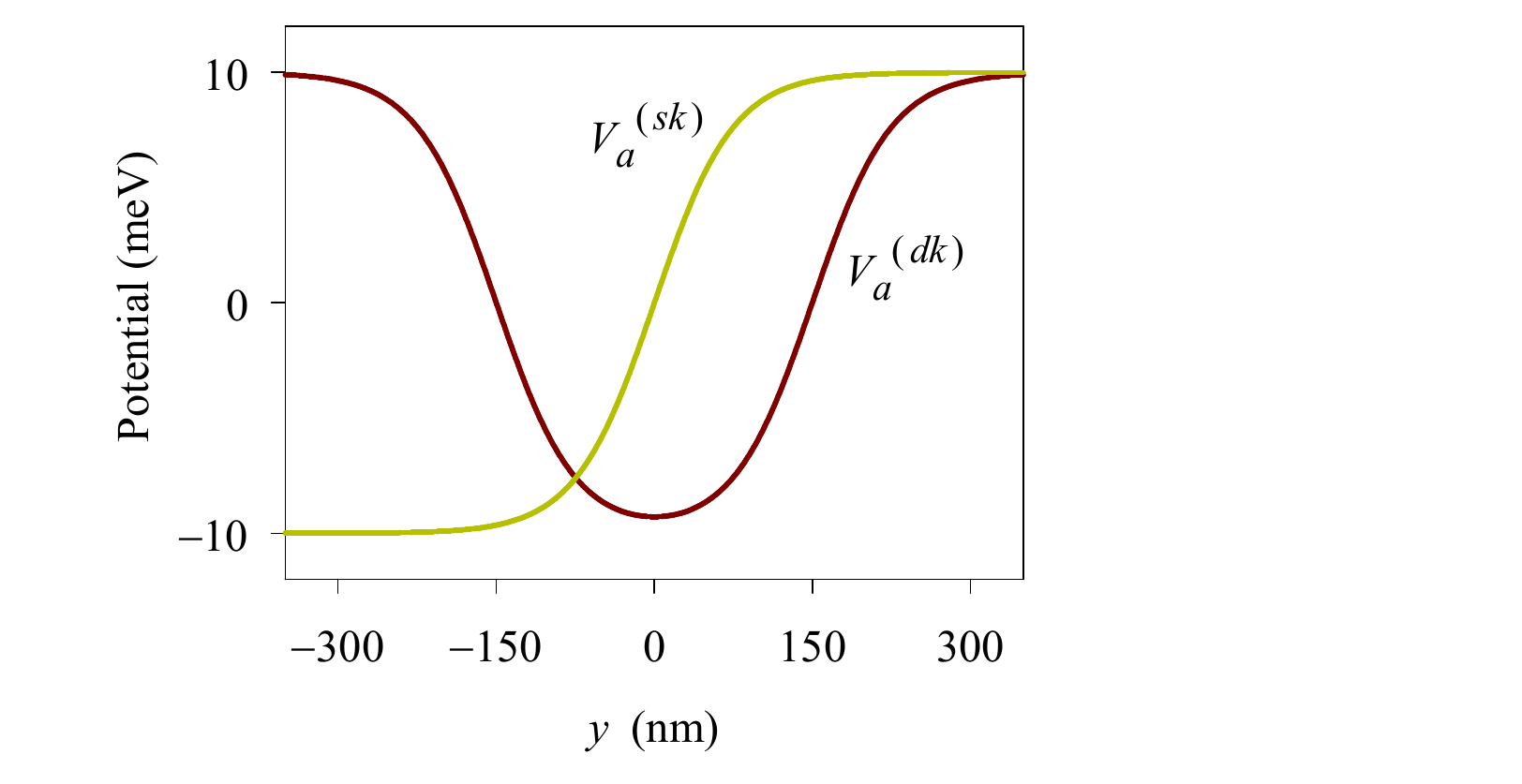}
\end{center}
\caption{ 
Smooth asymmetric potentials in a single kink and
a double kink (also called kink-antikink in the main text). 
Parameters: $V_a=10\,{\rm meV}$, $s=38\,{\rm nm}$, single kink $y_1=0$,
double kink $y_1=-150\, {\rm nm}$ and $y_2=150\, {\rm nm}$.
}
\label{F1S}
\end{figure}

Smoothness in the space variation of the asymmetric potential $V_a(y)$ 
is described with a logistic function. A smooth step at position $y_1$ and
diffusivity  $s$ (equivalently, the steepness inverse) is represented by 
\begin{equation}
{\cal F}(y,y_1,s)=\frac{1}{1+
e^{(y-y_1)/s}
}\; .
\end{equation}
In detail, the case of a single kink at $y_1$ reads 
\begin{eqnarray}
V_a^{(sk)}(y) &=&
V_a\, \left[1-2{\cal F}(y,y_1,s)\right]\; ,
\end{eqnarray}
with the asymptotic values
$V_a^{(sk)}(\pm\infty)=\pm V_a$.
In a straightforward extension, the double kink
forming a kink-antikink system
at $y_1$ and $y_2$ reads 
\begin{eqnarray}
V_a^{(dk)}(y) &=&
V_a\, \left[
1+2{\cal F}(y,y_1,s)-2{\cal F}(y,y_2,s)
\right]\; ,\nonumber\\
\end{eqnarray} 
Plots of single and double kinks with the above parametrizations 
are shown in Fig.\ \ref{F1S}.

\begin{figure}[t]
\begin{center}
\includegraphics[width=0.475\textwidth,trim=1cm 14.5cm 1.5cm 5.2cm,clip]{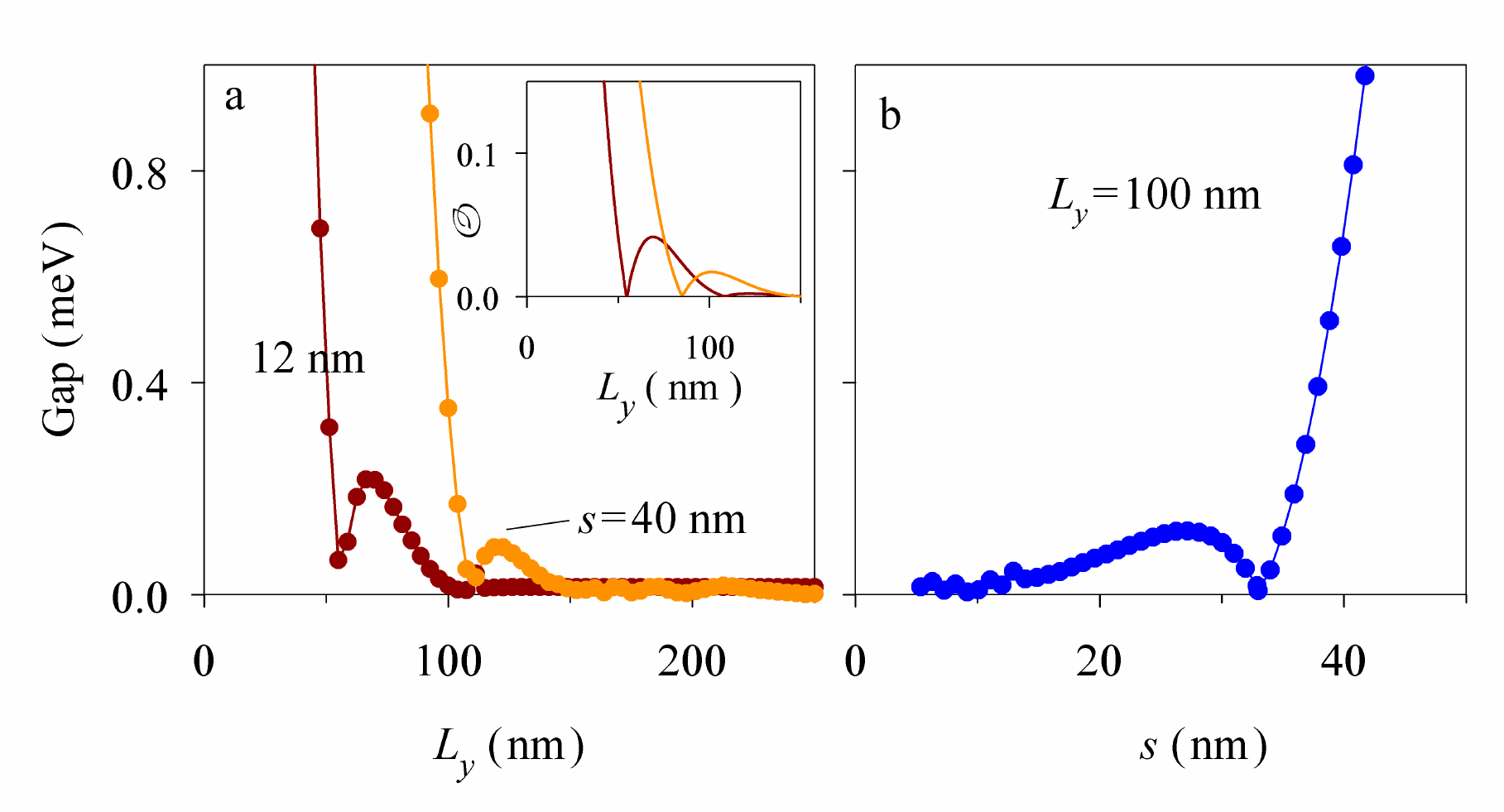}
\end{center}
\caption{
a) Energy gap of the kink-antikink band structure as a function of the 
separation $L_y$ and for two potential diffusivities $s$. The data points are obtained from the numerical band structure while the 
joining lines are a guide to the eye. The inset shows the wave function overlap ${\cal O}$ of independent kink and antinkink as a function of distance
with the same color code. b) Energy gap as a function of the diffusivity for a fixed separation.
Parameters: $V_a=10\,{\rm meV}$, $V_s=0$. 
}
\label{F2S}
\end{figure}

\section{Kink-antikink gap}\label{sec_gap}

 We address here how the gap of the kink-antinkink band structure  
(see Fig.\ \ref{Fig2}b) varies with the separation $L_y$ and the potential diffusivity $s$. The results are shown in Fig.\  \ref{F2S}a and b, respectively, and they confirm,
as was anticipated, that the gap strongly increases when $L_y$ decreases. However, it is remarkable that this dependence is non monotonic, with oscillations and with particular
values of $L_y$ and $s$ for which the gap vanishes. This 
behavior can be attributed to the oscillations of the wave functions~\cite{Mar08,Zarenia11}, as can be seen from the overlap of two displaced kink states
\begin{equation}
 {\cal O}(L_y)=
 \left|
 \int{dy\;
{\boldsymbol \phi}^{T*}(y)\,
{\boldsymbol \phi}(y-L_y)
}
\right|\; ,
\label{eqover}
\end{equation}
shown in the inset to Fig.\ \ref{F2S}a. In Eq.\ (\ref{eqover})
${\boldsymbol\phi}$ is the eight-component wave function
(and ${\boldsymbol\phi}^T$ its transpose)
for a real $k$ near the branch crossing at zero energy
of the single kink. For the steep potential in Fig.\ \ref{F2S} 
($s=12\,{\rm nm}$) the $L_y$ of minimum gap and the $L_y$ of 
vanishing overlap are  
in good agreement, while 
for the smooth potential ($s=38\,{\rm nm}$) the agreement is only qualitative.

\section{Scattering with complex band structure}\label{CBS}
Having mentioned in Sec.\ \ref{model} the overall idea of the complex-band-structure 
approach, we here present the details of the method. 
The double junction, Fig.\ \ref{Fsk}, is formed by three regions
$a=L,C,R$, each one homogenous along $x$. Introducing a local wave number $p_x\to \hbar k^{(a)}$ and a factorizing wave function $\phi_k^{(a)}(y\eta_\sigma\eta_\tau\eta_\lambda)e^{ik^{(a)}x}$,
with $\eta_{\sigma\tau\lambda}=1,2$ representing the pseudospin discrete components,
the energy eigenmode equation from 
Hamiltonian given by Eq.~(\ref{eq1}) for each region can be recast as
\begin{eqnarray}
& \left[
\frac{E}{v_F}\,\tau_z\sigma_x
+\hbar\frac{y}{l_z^2}
-i p_y \tau_z\sigma_z
-\frac{t}{2v_F}\,\tau_z(\lambda_x+i\lambda_y\sigma_z)
\right. & \nonumber\\
& \left.
-\frac{V_s}{v_F}\,\tau_z\sigma_x
-\frac{V_a(y)}{v_F}\,\lambda_z\tau_z\sigma_x
\right]
\phi_{k}^{(a)}
=\hbar k^{(a)}\, \phi_{k}^{(a)}\; ,
&
\label{eq1n}
\end{eqnarray}
where we left-multiplied all terms
by $\sigma_x\tau_z/v_F$, included a global sign change
and moved the $k^{(a)}$ linear term to the right-hand side.

Equation (\ref{eq1n}) is an eigenvalue equation for 
the mode sets $\{k^{(a)},\phi_k^{(a)}\}$ corresponding to a given energy $E$.
The problem is non-Hermitian due to the 
$ip_y\tau_z\sigma_z$ and $i\lambda_y\sigma_z$
contributions, physically allowing the possibility of complex wave numbers $k^{(a)}$. 
We solve Eq.\ (\ref{eq1n}) as a matrix eigenvalue problem, introducing a uniformly spaced 1D grid 
for the $y$ coordinate, determining the topological modes smoothly vanishing at the $y$ boundaries
with 
numerical methods well suited to large sparse matrices \cite{arpack}. 
The local complex wavenumbers 
and modes are our complex band structure basis sets in term of which we
represent the solution in each region of the double junction.

The wave function is given by a set of amplitudes $\{C_k^{(a)}\}$ 
in each $x$-homogenous region as
\begin{equation}
\Psi^{(a)}(xy\eta_\sigma\eta_\tau\eta_\lambda)=
\sum_{k^{(a)}}{
C_k^{(a)}\, 
\phi_k^{(a)}(y\eta_\sigma\eta_\tau\eta_\lambda)\,
e^{i k^{(a)}(x-x_k^{(a)})}\; .
}
\label{eq3n}
\end{equation}
The $x_k^{(a)}$ values in Eq.\ (\ref{eq3n}) are a gauge-like 
choice for the amplitudes $C_k^{(a)}$. This choice has 
its importance to avoid numerical instabilities due to exponentially
large numbers with complex wave numbers. For $a=L (R)$, the sum in Eq.\ (\ref{eq3n}) only includes 
wavenumbers with either zero or negative (positive) imaginary parts and we set  $x_k^{(L,R)}=x_{1,2}$, with $x_{1,2}$ the positions of the two 
interfaces. In the central region, however, Eq.\ (\ref{eq3n}) includes wave numbers with both $\leq 0$  and $>0$ imaginary parts, and we set $x_k^{(C)}=x_2$ and $x_1$ for those two cases, respectively.

At the two interfaces $x_{1,2}$ the wave function fulfills continuity.
This is the only requirement with a Dirac-like Hamiltonian like Eq.\ (\ref{eq1}), as opposed to a Schr\"odinger problem 
where the kinetic term also requires continuity of the wave function
first derivative. The continuity conditions yield the amplitudes
of outgoing modes $C_k^{(L,C,R)}$  in terms of those for input modes.
The latter are
assumed nonvanishing in the left lead only. 
Thus, we obtain a closed set of linear equations by projecting the continuity conditions at the two interfaces on the total set of
outgoing complex modes. 
In detail, the linear system reads
\begin{widetext}
\begin{equation}
\left\{
\begin{array}{rrll}
\displaystyle\sum_{k^{(L)}\,{\rm out}}{
{\cal M}_{k'k}^{(a L)}\,
{C_k^{(L)}}
}
&
-
\displaystyle\sum_{k^{(C)}\,{\rm out}}{
{\cal M}_{k'k}^{(a C)}\,
e^{i k^{(C)}(x_1-x_k^{(C)})}\,
{C_k^{(C)}}
}
&=
-\displaystyle\sum_{k^{(L)}\,{\rm in}\,{\rm prop.}}{
{\cal M}_{k'k}^{(a L)}\,
{C_k^{(L)}}
}\; ,
&
\quad {\rm if}\; x_{k'}^{(a)}=x_1\; , \\
\rule{0cm}{0.8cm}
\displaystyle\sum_{k^{(R)\,{\rm out}}}{
{\cal M}_{k'k}^{(a R)}\,
{C_k^{(R)}}
}
&
-
\displaystyle\sum_{k^{(C)}\,{\rm out}}{
{\cal M}_{k'k}^{(a C)}\,
e^{i k^{(C)}(x_2-x_k^{(C)})}\,
{C_k^{(C)}}
}
&
=0\; ,
&
\quad{\rm if}\; x_{k'}^{(a)}=x_2\; ,
\end{array}
\right.
\label{eq4n}
\end{equation}
\end{widetext}
where we have defined the overlap matrices
\begin{equation}
{\cal M}_{k'k}^{(a b)}
=
\sum_{\eta_\sigma\eta_\tau\eta_\lambda}
\int{ dy\,
\phi_{k'}^{(a) *}
\phi_{k}^{(b)}\; ,
}
\label{eq5n}
\end{equation}
and, as mentioned,  the sum on the right-hand side of Eq.\ (\ref{eq4n}) is restricted to the $k^{(L)}$ input propagating modes (with vanishing imaginary wavenumber) 
while those on the left-hand side contain output modes of all types, i.e., propagating
and evanescent.

The input/output character of the modes is summarized in 
Tab.\ \ref{tab1}.
Notice that 
in the central region all modes are output modes, while in left and right regions there can be input modes of propagating or evanescent character,
in our notation. In $L$ and $R$, the propagating modes  are classified as input or output depending on their quasiparticle flux, defined as
\begin{equation}
I_k^{(a)} = 
\langle \phi_k^{(a)} | v_F \tau_z\sigma_x | \phi_k^{(a)}\rangle\; ,
\label{eqC5}
\end{equation}
where $\partial H/\partial p_x = v_F \tau_z\sigma_x$ is the $x$-velocity operator.

Setting $C_k^{(L)}=1$ for a particular input propagating mode in Eq.\ (\ref{eq4n}),
with all other inputs vanishing, we obtain  a particular 
input/output pair transmission  $t_{n'n}\propto C_{k'}^{(L,R)}$ 
by solving the linear system.  We have numerically checked that
flux conservation is fulfilled in our method when the number of 
modes in each region is large enough. Typically, with 300 modes
in each region
we obtain a flux conservation with an error of 0.1{\%} 
or even smaller.

\begin{table}[b]
\begin{tabular}{c|c|c}
\hline\hline
region & mode character & condition \\
\hline
$C$    & output & none \\
$L$    & input/output prop.\ &$I_k^{(a)}>0 (<0)$; ${\rm Im}(k^{(a)})=0$\\
$L$    & input/output evan.\ & ${\rm Im}(k^{(a)})>0 (<0)$\\
$R$    & input/output prop.\ &$I_k^{(a)}<0 (>0)$; ${\rm Im}(k^{(a)})=0$\\
$R$    & input/output evan.\ & ${\rm Im}(k^{(a)})<0 (>0)$ \\
\hline\hline
\end{tabular}
\caption{Summary of the conditions for 
mode clasification as input or output in the three device regions.
$I_k^{(a)}$ is the mode flux defined in Eq.\ (\ref{eqC5}).
}
\label{tab1}
\end{table}

\bibliography{topval_prb_r2} 

\begin{thebibliography}{43}%
\makeatletter
\providecommand \@ifxundefined [1]{%
 \@ifx{#1\undefined}
}%
\providecommand \@ifnum [1]{%
 \ifnum #1\expandafter \@firstoftwo
 \else \expandafter \@secondoftwo
 \fi
}%
\providecommand \@ifx [1]{%
 \ifx #1\expandafter \@firstoftwo
 \else \expandafter \@secondoftwo
 \fi
}%
\providecommand \natexlab [1]{#1}%
\providecommand \enquote  [1]{``#1''}%
\providecommand \bibnamefont  [1]{#1}%
\providecommand \bibfnamefont [1]{#1}%
\providecommand \citenamefont [1]{#1}%
\providecommand \href@noop [0]{\@secondoftwo}%
\providecommand \href [0]{\begingroup \@sanitize@url \@href}%
\providecommand \@href[1]{\@@startlink{#1}\@@href}%
\providecommand \@@href[1]{\endgroup#1\@@endlink}%
\providecommand \@sanitize@url [0]{\catcode `\\12\catcode `\$12\catcode
  `\&12\catcode `\#12\catcode `\^12\catcode `\_12\catcode `\%12\relax}%
\providecommand \@@startlink[1]{}%
\providecommand \@@endlink[0]{}%
\providecommand \url  [0]{\begingroup\@sanitize@url \@url }%
\providecommand \@url [1]{\endgroup\@href {#1}{\urlprefix }}%
\providecommand \urlprefix  [0]{URL }%
\providecommand \Eprint [0]{\href }%
\providecommand \doibase [0]{http://dx.doi.org/}%
\providecommand \selectlanguage [0]{\@gobble}%
\providecommand \bibinfo  [0]{\@secondoftwo}%
\providecommand \bibfield  [0]{\@secondoftwo}%
\providecommand \translation [1]{[#1]}%
\providecommand \BibitemOpen [0]{}%
\providecommand \bibitemStop [0]{}%
\providecommand \bibitemNoStop [0]{.\EOS\space}%
\providecommand \EOS [0]{\spacefactor3000\relax}%
\providecommand \BibitemShut  [1]{\csname bibitem#1\endcsname}%
\let\auto@bib@innerbib\@empty
\bibitem [{\citenamefont {Trauzettel}\ \emph {et~al.}(2007)\citenamefont
  {Trauzettel}, \citenamefont {Bulaev}, \citenamefont {Loss},\ and\
  \citenamefont {Burkard}}]{Trau07}%
  \BibitemOpen
  \bibfield  {author} {\bibinfo {author} {\bibfnamefont {Bj{\"o}rn}\
  \bibnamefont {Trauzettel}}, \bibinfo {author} {\bibfnamefont {Denis~V.}\
  \bibnamefont {Bulaev}}, \bibinfo {author} {\bibfnamefont {Daniel}\
  \bibnamefont {Loss}}, \ and\ \bibinfo {author} {\bibfnamefont {Guido}\
  \bibnamefont {Burkard}},\ }\bibfield  {title} {\enquote {\bibinfo {title}
  {Spin qubits in graphene quantum dots},}\ }\href@noop {} {\bibfield
  {journal} {\bibinfo  {journal} {Nature Physics}\ }\textbf {\bibinfo {volume}
  {3}},\ \bibinfo {pages} {192--196} (\bibinfo {year} {2007})}\BibitemShut
  {NoStop}%
\bibitem [{\citenamefont {Pereira}\ \emph {et~al.}(2007)\citenamefont
  {Pereira}, \citenamefont {Vasilopoulos},\ and\ \citenamefont
  {Peeters}}]{Pereira07}%
  \BibitemOpen
  \bibfield  {author} {\bibinfo {author} {\bibfnamefont {J.~Milton}\
  \bibnamefont {Pereira}}, \bibinfo {author} {\bibfnamefont {P.}~\bibnamefont
  {Vasilopoulos}}, \ and\ \bibinfo {author} {\bibfnamefont {F.~M.}\
  \bibnamefont {Peeters}},\ }\bibfield  {title} {\enquote {\bibinfo {title}
  {Tunable quantum dots in bilayer graphene},}\ }\href {\doibase
  10.1021/nl062967s} {\bibfield  {journal} {\bibinfo  {journal} {Nano Letters}\
  }\textbf {\bibinfo {volume} {7}},\ \bibinfo {pages} {946--949} (\bibinfo
  {year} {2007})}\BibitemShut {NoStop}%
\bibitem [{\citenamefont {Rycerz}\ \emph {et~al.}(2007)\citenamefont {Rycerz},
  \citenamefont {Tworzyd{\l}o},\ and\ \citenamefont {Beenakker}}]{Ryc07}%
  \BibitemOpen
  \bibfield  {author} {\bibinfo {author} {\bibfnamefont {A.}~\bibnamefont
  {Rycerz}}, \bibinfo {author} {\bibfnamefont {J.}~\bibnamefont
  {Tworzyd{\l}o}}, \ and\ \bibinfo {author} {\bibfnamefont {C.~W.~J.}\
  \bibnamefont {Beenakker}},\ }\bibfield  {title} {\enquote {\bibinfo {title}
  {Valley filter and valley valve in graphene},}\ }\href {\doibase
  10.1038/nphys547} {\bibfield  {journal} {\bibinfo  {journal} {Nature
  Physics}\ }\textbf {\bibinfo {volume} {3}},\ \bibinfo {pages} {172--175}
  (\bibinfo {year} {2007})}\BibitemShut {NoStop}%
\bibitem [{\citenamefont {Recher}\ and\ \citenamefont
  {Trauzettel}(2010)}]{Recher10}%
  \BibitemOpen
  \bibfield  {author} {\bibinfo {author} {\bibfnamefont {Patrik}\ \bibnamefont
  {Recher}}\ and\ \bibinfo {author} {\bibfnamefont {Bj{\"o}rn}\ \bibnamefont
  {Trauzettel}},\ }\bibfield  {title} {\enquote {\bibinfo {title} {Quantum dots
  and spin qubits in graphene},}\ }\href {\doibase
  10.1088/0957-4484/21/30/302001} {\bibfield  {journal} {\bibinfo  {journal}
  {Nanotechnology}\ }\textbf {\bibinfo {volume} {21}},\ \bibinfo {pages}
  {302001} (\bibinfo {year} {2010})}\BibitemShut {NoStop}%
\bibitem [{\citenamefont {Gunlycke}\ and\ \citenamefont {White}(2011)}]{Gun11}%
  \BibitemOpen
  \bibfield  {author} {\bibinfo {author} {\bibfnamefont {D.}~\bibnamefont
  {Gunlycke}}\ and\ \bibinfo {author} {\bibfnamefont {C.~T.}\ \bibnamefont
  {White}},\ }\bibfield  {title} {\enquote {\bibinfo {title} {Graphene valley
  filter using a line defect},}\ }\href {\doibase
  10.1103/PhysRevLett.106.136806} {\bibfield  {journal} {\bibinfo  {journal}
  {Phys. Rev. Lett.}\ }\textbf {\bibinfo {volume} {106}},\ \bibinfo {pages}
  {136806} (\bibinfo {year} {2011})}\BibitemShut {NoStop}%
\bibitem [{\citenamefont {McCann}\ and\ \citenamefont
  {Koshino}(2013)}]{Mcan13}%
  \BibitemOpen
  \bibfield  {author} {\bibinfo {author} {\bibfnamefont {Edward}\ \bibnamefont
  {McCann}}\ and\ \bibinfo {author} {\bibfnamefont {Mikito}\ \bibnamefont
  {Koshino}},\ }\bibfield  {title} {\enquote {\bibinfo {title} {The electronic
  properties of bilayer graphene},}\ }\href {\doibase
  10.1088/0034-4885/76/5/056503} {\bibfield  {journal} {\bibinfo  {journal}
  {Reports on Progress in Physics}\ }\textbf {\bibinfo {volume} {76}},\
  \bibinfo {pages} {056503} (\bibinfo {year} {2013})}\BibitemShut {NoStop}%
\bibitem [{\citenamefont {Rozhkov}\ \emph {et~al.}(2016)\citenamefont
  {Rozhkov}, \citenamefont {Sboychakov}, \citenamefont {Rakhmanov},\ and\
  \citenamefont {Nori}}]{rozhkov16}%
  \BibitemOpen
  \bibfield  {author} {\bibinfo {author} {\bibfnamefont {A.V.}\ \bibnamefont
  {Rozhkov}}, \bibinfo {author} {\bibfnamefont {A.O.}\ \bibnamefont
  {Sboychakov}}, \bibinfo {author} {\bibfnamefont {A.L.}\ \bibnamefont
  {Rakhmanov}}, \ and\ \bibinfo {author} {\bibfnamefont {Franco}\ \bibnamefont
  {Nori}},\ }\bibfield  {title} {\enquote {\bibinfo {title} {Electronic
  properties of graphene-based bilayer systems},}\ }\href {\doibase
  https://doi.org/10.1016/j.physrep.2016.07.003} {\bibfield  {journal}
  {\bibinfo  {journal} {Physics Reports}\ }\textbf {\bibinfo {volume} {648}},\
  \bibinfo {pages} {1--104} (\bibinfo {year} {2016})},\ \bibinfo {note}
  {electronic properties of graphene-based bilayer systems}\BibitemShut
  {NoStop}%
\bibitem [{\citenamefont {Zhang}\ \emph {et~al.}(2009)\citenamefont {Zhang},
  \citenamefont {Tang}, \citenamefont {Girit}, \citenamefont {Hao},
  \citenamefont {Martin}, \citenamefont {Zettl}, \citenamefont {Crommie},
  \citenamefont {Shen},\ and\ \citenamefont {Wang}}]{Zhang09}%
  \BibitemOpen
  \bibfield  {author} {\bibinfo {author} {\bibfnamefont {Y.}~\bibnamefont
  {Zhang}}, \bibinfo {author} {\bibfnamefont {T.-T.}\ \bibnamefont {Tang}},
  \bibinfo {author} {\bibfnamefont {C.}~\bibnamefont {Girit}}, \bibinfo
  {author} {\bibfnamefont {Z.}~\bibnamefont {Hao}}, \bibinfo {author}
  {\bibfnamefont {M.~C.}\ \bibnamefont {Martin}}, \bibinfo {author}
  {\bibfnamefont {A.}~\bibnamefont {Zettl}}, \bibinfo {author} {\bibfnamefont
  {M.~F.}\ \bibnamefont {Crommie}}, \bibinfo {author} {\bibfnamefont {Y.~R.}\
  \bibnamefont {Shen}}, \ and\ \bibinfo {author} {\bibfnamefont
  {F.}~\bibnamefont {Wang}},\ }\bibfield  {title} {\enquote {\bibinfo {title}
  {Direct observation of a widely tunable bandgap in bilayer graphene},}\
  }\href {\doibase 10.1038/nature08105} {\bibfield  {journal} {\bibinfo
  {journal} {Nature}\ }\textbf {\bibinfo {volume} {459}},\ \bibinfo {pages}
  {820--823} (\bibinfo {year} {2009})}\BibitemShut {NoStop}%
\bibitem [{\citenamefont {Overweg}\ \emph
  {et~al.}(2018{\natexlab{a}})\citenamefont {Overweg}, \citenamefont
  {Eggimann}, \citenamefont {Chen}, \citenamefont {Slizovskiy}, \citenamefont
  {Eich}, \citenamefont {Pisoni}, \citenamefont {Lee}, \citenamefont
  {Rickhaus}, \citenamefont {Watanabe}, \citenamefont {Taniguchi},
  \citenamefont {Fal{'k}o}, \citenamefont {Ihn},\ and\ \citenamefont
  {Ensslin}}]{Overweg18}%
  \BibitemOpen
  \bibfield  {author} {\bibinfo {author} {\bibfnamefont {Hiske}\ \bibnamefont
  {Overweg}}, \bibinfo {author} {\bibfnamefont {Hannah}\ \bibnamefont
  {Eggimann}}, \bibinfo {author} {\bibfnamefont {Xi}~\bibnamefont {Chen}},
  \bibinfo {author} {\bibfnamefont {Sergey}\ \bibnamefont {Slizovskiy}},
  \bibinfo {author} {\bibfnamefont {Marius}\ \bibnamefont {Eich}}, \bibinfo
  {author} {\bibfnamefont {Riccardo}\ \bibnamefont {Pisoni}}, \bibinfo {author}
  {\bibfnamefont {Yongjin}\ \bibnamefont {Lee}}, \bibinfo {author}
  {\bibfnamefont {Peter}\ \bibnamefont {Rickhaus}}, \bibinfo {author}
  {\bibfnamefont {Kenji}\ \bibnamefont {Watanabe}}, \bibinfo {author}
  {\bibfnamefont {Takashi}\ \bibnamefont {Taniguchi}}, \bibinfo {author}
  {\bibfnamefont {Vladimir}\ \bibnamefont {Fal{'k}o}}, \bibinfo {author}
  {\bibfnamefont {Thomas}\ \bibnamefont {Ihn}}, \ and\ \bibinfo {author}
  {\bibfnamefont {Klaus}\ \bibnamefont {Ensslin}},\ }\bibfield  {title}
  {\enquote {\bibinfo {title} {Electrostatically induced quantum point contacts
  in bilayer graphene},}\ }\href {\doibase 10.1021/acs.nanolett.7b04666}
  {\bibfield  {journal} {\bibinfo  {journal} {Nano Letters}\ }\textbf {\bibinfo
  {volume} {18}},\ \bibinfo {pages} {553--559} (\bibinfo {year}
  {2018}{\natexlab{a}})}\BibitemShut {NoStop}%
\bibitem [{\citenamefont {Overweg}\ \emph
  {et~al.}(2018{\natexlab{b}})\citenamefont {Overweg}, \citenamefont {Knothe},
  \citenamefont {Fabian}, \citenamefont {Linhart}, \citenamefont {Rickhaus},
  \citenamefont {Wernli}, \citenamefont {Watanabe}, \citenamefont {Taniguchi},
  \citenamefont {S\'anchez}, \citenamefont {Burgd\"orfer}, \citenamefont
  {Libisch}, \citenamefont {Fal{'k}o}, \citenamefont {Ensslin},\ and\
  \citenamefont {Ihn}}]{Over18}%
  \BibitemOpen
  \bibfield  {author} {\bibinfo {author} {\bibfnamefont {Hiske}\ \bibnamefont
  {Overweg}}, \bibinfo {author} {\bibfnamefont {Angelika}\ \bibnamefont
  {Knothe}}, \bibinfo {author} {\bibfnamefont {Thomas}\ \bibnamefont {Fabian}},
  \bibinfo {author} {\bibfnamefont {Lukas}\ \bibnamefont {Linhart}}, \bibinfo
  {author} {\bibfnamefont {Peter}\ \bibnamefont {Rickhaus}}, \bibinfo {author}
  {\bibfnamefont {Lucien}\ \bibnamefont {Wernli}}, \bibinfo {author}
  {\bibfnamefont {Kenji}\ \bibnamefont {Watanabe}}, \bibinfo {author}
  {\bibfnamefont {Takashi}\ \bibnamefont {Taniguchi}}, \bibinfo {author}
  {\bibfnamefont {David}\ \bibnamefont {S\'anchez}}, \bibinfo {author}
  {\bibfnamefont {Joachim}\ \bibnamefont {Burgd\"orfer}}, \bibinfo {author}
  {\bibfnamefont {Florian}\ \bibnamefont {Libisch}}, \bibinfo {author}
  {\bibfnamefont {Vladimir~I.}\ \bibnamefont {Fal{'k}o}}, \bibinfo {author}
  {\bibfnamefont {Klaus}\ \bibnamefont {Ensslin}}, \ and\ \bibinfo {author}
  {\bibfnamefont {Thomas}\ \bibnamefont {Ihn}},\ }\bibfield  {title} {\enquote
  {\bibinfo {title} {Topologically nontrivial valley states in bilayer graphene
  quantum point contacts},}\ }\href {\doibase 10.1103/PhysRevLett.121.257702}
  {\bibfield  {journal} {\bibinfo  {journal} {Phys. Rev. Lett.}\ }\textbf
  {\bibinfo {volume} {121}},\ \bibinfo {pages} {257702} (\bibinfo {year}
  {2018}{\natexlab{b}})}\BibitemShut {NoStop}%
\bibitem [{\citenamefont {Kraft}\ \emph {et~al.}(2018)\citenamefont {Kraft},
  \citenamefont {Krainov}, \citenamefont {Gall}, \citenamefont {Dmitriev},
  \citenamefont {Krupke}, \citenamefont {Gornyi},\ and\ \citenamefont
  {Danneau}}]{Kraf18}%
  \BibitemOpen
  \bibfield  {author} {\bibinfo {author} {\bibfnamefont {R.}~\bibnamefont
  {Kraft}}, \bibinfo {author} {\bibfnamefont {I.~V.}\ \bibnamefont {Krainov}},
  \bibinfo {author} {\bibfnamefont {V.}~\bibnamefont {Gall}}, \bibinfo {author}
  {\bibfnamefont {A.~P.}\ \bibnamefont {Dmitriev}}, \bibinfo {author}
  {\bibfnamefont {R.}~\bibnamefont {Krupke}}, \bibinfo {author} {\bibfnamefont
  {I.~V.}\ \bibnamefont {Gornyi}}, \ and\ \bibinfo {author} {\bibfnamefont
  {R.}~\bibnamefont {Danneau}},\ }\bibfield  {title} {\enquote {\bibinfo
  {title} {Valley subband splitting in bilayer graphene quantum point
  contacts},}\ }\href {\doibase 10.1103/PhysRevLett.121.257703} {\bibfield
  {journal} {\bibinfo  {journal} {Phys. Rev. Lett.}\ }\textbf {\bibinfo
  {volume} {121}},\ \bibinfo {pages} {257703} (\bibinfo {year}
  {2018})}\BibitemShut {NoStop}%
\bibitem [{\citenamefont {Terr\'es}\ \emph {et~al.}(2016)\citenamefont
  {Terr\'es}, \citenamefont {Chizhova}, \citenamefont {Libisch}, \citenamefont
  {Peiro}, \citenamefont {J\"orger}, \citenamefont {Engels}, \citenamefont
  {Girschik}, \citenamefont {Watanabe}, \citenamefont {Taniguchi},
  \citenamefont {Rotking},\ and\ \citenamefont {Stampfer}}]{Ter16}%
  \BibitemOpen
  \bibfield  {author} {\bibinfo {author} {\bibfnamefont {B.}~\bibnamefont
  {Terr\'es}}, \bibinfo {author} {\bibfnamefont {L.~A.}\ \bibnamefont
  {Chizhova}}, \bibinfo {author} {\bibfnamefont {F.}~\bibnamefont {Libisch}},
  \bibinfo {author} {\bibfnamefont {J.}~\bibnamefont {Peiro}}, \bibinfo
  {author} {\bibfnamefont {D.}~\bibnamefont {J\"orger}}, \bibinfo {author}
  {\bibfnamefont {S.}~\bibnamefont {Engels}}, \bibinfo {author} {\bibfnamefont
  {A.}~\bibnamefont {Girschik}}, \bibinfo {author} {\bibfnamefont
  {K.}~\bibnamefont {Watanabe}}, \bibinfo {author} {\bibfnamefont
  {T.}~\bibnamefont {Taniguchi}}, \bibinfo {author} {\bibfnamefont {S.~V.}\
  \bibnamefont {Rotking}}, \ and\ \bibinfo {author} {\bibfnamefont
  {C.}~\bibnamefont {Stampfer}},\ }\bibfield  {title} {\enquote {\bibinfo
  {title} {Size quantization of dirac fermions in graphene constrictions},}\
  }\href@noop {} {\bibfield  {journal} {\bibinfo  {journal} {Nature
  Communications}\ }\textbf {\bibinfo {volume} {7}},\ \bibinfo {pages} {11528}
  (\bibinfo {year} {2016})}\BibitemShut {NoStop}%
\bibitem [{\citenamefont {Cleric\`o}\ \emph {et~al.}(2019)\citenamefont
  {Cleric\`o}, \citenamefont {Delgado-Notario}, \citenamefont
  {Saiz-Bret\'{\i}n}, \citenamefont {Malyshev}, \citenamefont {Meziani},
  \citenamefont {Hidalgo}, \citenamefont {M\'endez}, \citenamefont {Amado},
  \citenamefont {Dom\'{\i}nguez-Adame},\ and\ \citenamefont {Diez}}]{Cle19}%
  \BibitemOpen
  \bibfield  {author} {\bibinfo {author} {\bibfnamefont {V.}~\bibnamefont
  {Cleric\`o}}, \bibinfo {author} {\bibfnamefont {J.~A.}\ \bibnamefont
  {Delgado-Notario}}, \bibinfo {author} {\bibfnamefont {M.}~\bibnamefont
  {Saiz-Bret\'{\i}n}}, \bibinfo {author} {\bibfnamefont {A.~V.}\ \bibnamefont
  {Malyshev}}, \bibinfo {author} {\bibfnamefont {Y.~M.}\ \bibnamefont
  {Meziani}}, \bibinfo {author} {\bibfnamefont {P.}~\bibnamefont {Hidalgo}},
  \bibinfo {author} {\bibfnamefont {B.}~\bibnamefont {M\'endez}}, \bibinfo
  {author} {\bibfnamefont {M.}~\bibnamefont {Amado}}, \bibinfo {author}
  {\bibfnamefont {F.}~\bibnamefont {Dom\'{\i}nguez-Adame}}, \ and\ \bibinfo
  {author} {\bibfnamefont {E.}~\bibnamefont {Diez}},\ }\bibfield  {title}
  {\enquote {\bibinfo {title} {Quantum nanoconstrictions fabricated by
  cryo-etching in encapsulated graphene},}\ }\href@noop {} {\bibfield
  {journal} {\bibinfo  {journal} {Sci. Rep.}\ }\textbf {\bibinfo {volume}
  {9}},\ \bibinfo {pages} {13572} (\bibinfo {year} {2019})}\BibitemShut
  {NoStop}%
\bibitem [{\citenamefont {Eich}\ \emph {et~al.}(2018)\citenamefont {Eich},
  \citenamefont {Herman}, \citenamefont {Pisoni}, \citenamefont {Overweg},
  \citenamefont {Kurzmann}, \citenamefont {Lee}, \citenamefont {Rickhaus},
  \citenamefont {Watanabe}, \citenamefont {Taniguchi}, \citenamefont {Sigrist},
  \citenamefont {Ihn},\ and\ \citenamefont {Ensslin}}]{Eich18}%
  \BibitemOpen
  \bibfield  {author} {\bibinfo {author} {\bibfnamefont {Marius}\ \bibnamefont
  {Eich}}, \bibinfo {author} {\bibfnamefont {F.}~\bibnamefont {Herman}},
  \bibinfo {author} {\bibfnamefont {Riccardo}\ \bibnamefont {Pisoni}}, \bibinfo
  {author} {\bibfnamefont {Hiske}\ \bibnamefont {Overweg}}, \bibinfo {author}
  {\bibfnamefont {Annika}\ \bibnamefont {Kurzmann}}, \bibinfo {author}
  {\bibfnamefont {Yongjin}\ \bibnamefont {Lee}}, \bibinfo {author}
  {\bibfnamefont {Peter}\ \bibnamefont {Rickhaus}}, \bibinfo {author}
  {\bibfnamefont {Kenji}\ \bibnamefont {Watanabe}}, \bibinfo {author}
  {\bibfnamefont {Takashi}\ \bibnamefont {Taniguchi}}, \bibinfo {author}
  {\bibfnamefont {Manfred}\ \bibnamefont {Sigrist}}, \bibinfo {author}
  {\bibfnamefont {Thomas}\ \bibnamefont {Ihn}}, \ and\ \bibinfo {author}
  {\bibfnamefont {Klaus}\ \bibnamefont {Ensslin}},\ }\bibfield  {title}
  {\enquote {\bibinfo {title} {Spin and valley states in gate-defined bilayer
  graphene quantum dots},}\ }\href {\doibase 10.1103/PhysRevX.8.031023}
  {\bibfield  {journal} {\bibinfo  {journal} {Phys. Rev. X}\ }\textbf {\bibinfo
  {volume} {8}},\ \bibinfo {pages} {031023} (\bibinfo {year}
  {2018})}\BibitemShut {NoStop}%
\bibitem [{\citenamefont {Kurzmann}\ \emph {et~al.}(2019)\citenamefont
  {Kurzmann}, \citenamefont {Overweg}, \citenamefont {Eich}, \citenamefont
  {Pally}, \citenamefont {Rickhaus}, \citenamefont {Pisoni}, \citenamefont
  {Lee}, \citenamefont {Watanabe}, \citenamefont {Taniguchi}, \citenamefont
  {Ihn},\ and\ \citenamefont {Ensslin}}]{Kurzmann19}%
  \BibitemOpen
  \bibfield  {author} {\bibinfo {author} {\bibfnamefont {Annika}\ \bibnamefont
  {Kurzmann}}, \bibinfo {author} {\bibfnamefont {Hiske}\ \bibnamefont
  {Overweg}}, \bibinfo {author} {\bibfnamefont {Marius}\ \bibnamefont {Eich}},
  \bibinfo {author} {\bibfnamefont {Alessia}\ \bibnamefont {Pally}}, \bibinfo
  {author} {\bibfnamefont {Peter}\ \bibnamefont {Rickhaus}}, \bibinfo {author}
  {\bibfnamefont {Riccardo}\ \bibnamefont {Pisoni}}, \bibinfo {author}
  {\bibfnamefont {Yongjin}\ \bibnamefont {Lee}}, \bibinfo {author}
  {\bibfnamefont {Kenji}\ \bibnamefont {Watanabe}}, \bibinfo {author}
  {\bibfnamefont {Takashi}\ \bibnamefont {Taniguchi}}, \bibinfo {author}
  {\bibfnamefont {Thomas}\ \bibnamefont {Ihn}}, \ and\ \bibinfo {author}
  {\bibfnamefont {Klaus}\ \bibnamefont {Ensslin}},\ }\bibfield  {title}
  {\enquote {\bibinfo {title} {Charge detection in gate-defined bilayer
  graphene quantum dots},}\ }\href {\doibase 10.1021/acs.nanolett.9b01617}
  {\bibfield  {journal} {\bibinfo  {journal} {Nano Letters}\ }\textbf {\bibinfo
  {volume} {19}},\ \bibinfo {pages} {5216--5221} (\bibinfo {year}
  {2019})}\BibitemShut {NoStop}%
\bibitem [{\citenamefont {Banszerus}\ \emph {et~al.}(2020)\citenamefont
  {Banszerus}, \citenamefont {Rothstein}, \citenamefont {Fabian}, \citenamefont
  {M{\"o}ller}, \citenamefont {Icking}, \citenamefont {Trellenkamp},
  \citenamefont {Lentz}, \citenamefont {Neumaier}, \citenamefont {Watanabe},
  \citenamefont {Taniguchi}, \citenamefont {Libisch}, \citenamefont {Volk},\
  and\ \citenamefont {Stampfer}}]{Banszerus20}%
  \BibitemOpen
  \bibfield  {author} {\bibinfo {author} {\bibfnamefont {L.}~\bibnamefont
  {Banszerus}}, \bibinfo {author} {\bibfnamefont {A.}~\bibnamefont
  {Rothstein}}, \bibinfo {author} {\bibfnamefont {T.}~\bibnamefont {Fabian}},
  \bibinfo {author} {\bibfnamefont {S.}~\bibnamefont {M{\"o}ller}}, \bibinfo
  {author} {\bibfnamefont {E.}~\bibnamefont {Icking}}, \bibinfo {author}
  {\bibfnamefont {S.}~\bibnamefont {Trellenkamp}}, \bibinfo {author}
  {\bibfnamefont {F.}~\bibnamefont {Lentz}}, \bibinfo {author} {\bibfnamefont
  {D.}~\bibnamefont {Neumaier}}, \bibinfo {author} {\bibfnamefont
  {K.}~\bibnamefont {Watanabe}}, \bibinfo {author} {\bibfnamefont
  {T.}~\bibnamefont {Taniguchi}}, \bibinfo {author} {\bibfnamefont
  {F.}~\bibnamefont {Libisch}}, \bibinfo {author} {\bibfnamefont
  {C.}~\bibnamefont {Volk}}, \ and\ \bibinfo {author} {\bibfnamefont
  {C.}~\bibnamefont {Stampfer}},\ }\bibfield  {title} {\enquote {\bibinfo
  {title} {Electron hole crossover in gate-controlled bilayer graphene quantum
  dots},}\ }\href {\doibase 10.1021/acs.nanolett.0c03227} {\bibfield  {journal}
  {\bibinfo  {journal} {Nano Letters}\ }\textbf {\bibinfo {volume} {20}},\
  \bibinfo {pages} {7709--7715} (\bibinfo {year} {2020})}\BibitemShut {NoStop}%
\bibitem [{\citenamefont {Banszerus}\ \emph {et~al.}(2021)\citenamefont
  {Banszerus}, \citenamefont {Hecker}, \citenamefont {Icking}, \citenamefont
  {Trellenkamp}, \citenamefont {Lentz}, \citenamefont {Neumaier}, \citenamefont
  {Watanabe}, \citenamefont {Taniguchi}, \citenamefont {Volk},\ and\
  \citenamefont {Stampfer}}]{Banszerus21}%
  \BibitemOpen
  \bibfield  {author} {\bibinfo {author} {\bibfnamefont {L.}~\bibnamefont
  {Banszerus}}, \bibinfo {author} {\bibfnamefont {K.}~\bibnamefont {Hecker}},
  \bibinfo {author} {\bibfnamefont {E.}~\bibnamefont {Icking}}, \bibinfo
  {author} {\bibfnamefont {S.}~\bibnamefont {Trellenkamp}}, \bibinfo {author}
  {\bibfnamefont {F.}~\bibnamefont {Lentz}}, \bibinfo {author} {\bibfnamefont
  {D.}~\bibnamefont {Neumaier}}, \bibinfo {author} {\bibfnamefont
  {K.}~\bibnamefont {Watanabe}}, \bibinfo {author} {\bibfnamefont
  {T.}~\bibnamefont {Taniguchi}}, \bibinfo {author} {\bibfnamefont
  {C.}~\bibnamefont {Volk}}, \ and\ \bibinfo {author} {\bibfnamefont
  {C.}~\bibnamefont {Stampfer}},\ }\bibfield  {title} {\enquote {\bibinfo
  {title} {Pulsed-gate spectroscopy of single-electron spin states in bilayer
  graphene quantum dots},}\ }\href {\doibase 10.1103/PhysRevB.103.L081404}
  {\bibfield  {journal} {\bibinfo  {journal} {Phys. Rev. B}\ }\textbf {\bibinfo
  {volume} {103}},\ \bibinfo {pages} {L081404} (\bibinfo {year}
  {2021})}\BibitemShut {NoStop}%
\bibitem [{\citenamefont {Martin}\ \emph {et~al.}(2008)\citenamefont {Martin},
  \citenamefont {Blanter},\ and\ \citenamefont {Morpurgo}}]{Mar08}%
  \BibitemOpen
  \bibfield  {author} {\bibinfo {author} {\bibfnamefont {Ivar}\ \bibnamefont
  {Martin}}, \bibinfo {author} {\bibfnamefont {Ya.~M.}\ \bibnamefont
  {Blanter}}, \ and\ \bibinfo {author} {\bibfnamefont {A.~F.}\ \bibnamefont
  {Morpurgo}},\ }\bibfield  {title} {\enquote {\bibinfo {title} {Topological
  confinement in bilayer graphene},}\ }\href {\doibase
  10.1103/PhysRevLett.100.036804} {\bibfield  {journal} {\bibinfo  {journal}
  {Phys. Rev. Lett.}\ }\textbf {\bibinfo {volume} {100}},\ \bibinfo {pages}
  {036804} (\bibinfo {year} {2008})}\BibitemShut {NoStop}%
\bibitem [{\citenamefont {Zarenia}\ \emph {et~al.}(2011)\citenamefont
  {Zarenia}, \citenamefont {Pereira}, \citenamefont {Farias},\ and\
  \citenamefont {Peeters}}]{Zarenia11}%
  \BibitemOpen
  \bibfield  {author} {\bibinfo {author} {\bibfnamefont {M.}~\bibnamefont
  {Zarenia}}, \bibinfo {author} {\bibfnamefont {J.~M.}\ \bibnamefont
  {Pereira}}, \bibinfo {author} {\bibfnamefont {G.~A.}\ \bibnamefont {Farias}},
  \ and\ \bibinfo {author} {\bibfnamefont {F.~M.}\ \bibnamefont {Peeters}},\
  }\bibfield  {title} {\enquote {\bibinfo {title} {Chiral states in bilayer
  graphene: Magnetic field dependence and gap opening},}\ }\href {\doibase
  10.1103/PhysRevB.84.125451} {\bibfield  {journal} {\bibinfo  {journal} {Phys.
  Rev. B}\ }\textbf {\bibinfo {volume} {84}},\ \bibinfo {pages} {125451}
  (\bibinfo {year} {2011})}\BibitemShut {NoStop}%
\bibitem [{\citenamefont {Zhang}\ \emph {et~al.}(2013)\citenamefont {Zhang},
  \citenamefont {MacDonald},\ and\ \citenamefont {Mele}}]{Zhang13}%
  \BibitemOpen
  \bibfield  {author} {\bibinfo {author} {\bibfnamefont {Fan}\ \bibnamefont
  {Zhang}}, \bibinfo {author} {\bibfnamefont {Allan~H.}\ \bibnamefont
  {MacDonald}}, \ and\ \bibinfo {author} {\bibfnamefont {Eugene~J.}\
  \bibnamefont {Mele}},\ }\bibfield  {title} {\enquote {\bibinfo {title}
  {Valley chern numbers and boundary modes in gapped bilayer graphene},}\
  }\href {\doibase 10.1073/pnas.1308853110} {\bibfield  {journal} {\bibinfo
  {journal} {Proceedings of the National Academy of Sciences}\ }\textbf
  {\bibinfo {volume} {110}},\ \bibinfo {pages} {10546--10551} (\bibinfo {year}
  {2013})}\BibitemShut {NoStop}%
\bibitem [{\citenamefont {Ju}\ \emph {et~al.}(2015)\citenamefont {Ju},
  \citenamefont {Shi}, \citenamefont {Nair}, \citenamefont {Lv}, \citenamefont
  {Jin}, \citenamefont {Velasco}, \citenamefont {Ojeda-Aristizabal},
  \citenamefont {Bechtel}, \citenamefont {Martin}, \citenamefont {Zettl},
  \citenamefont {Analytis},\ and\ \citenamefont {Wang}}]{Lon15}%
  \BibitemOpen
  \bibfield  {author} {\bibinfo {author} {\bibfnamefont {Long}\ \bibnamefont
  {Ju}}, \bibinfo {author} {\bibfnamefont {Zhiwen}\ \bibnamefont {Shi}},
  \bibinfo {author} {\bibfnamefont {Nityan}\ \bibnamefont {Nair}}, \bibinfo
  {author} {\bibfnamefont {Yinchuan}\ \bibnamefont {Lv}}, \bibinfo {author}
  {\bibfnamefont {Chenhao}\ \bibnamefont {Jin}}, \bibinfo {author}
  {\bibfnamefont {Jairo}\ \bibnamefont {Velasco}}, \bibinfo {author}
  {\bibfnamefont {Claudia}\ \bibnamefont {Ojeda-Aristizabal}}, \bibinfo
  {author} {\bibfnamefont {Hans~A.}\ \bibnamefont {Bechtel}}, \bibinfo {author}
  {\bibfnamefont {Michael~C.}\ \bibnamefont {Martin}}, \bibinfo {author}
  {\bibfnamefont {Alex}\ \bibnamefont {Zettl}}, \bibinfo {author}
  {\bibfnamefont {James}\ \bibnamefont {Analytis}}, \ and\ \bibinfo {author}
  {\bibfnamefont {Feng}\ \bibnamefont {Wang}},\ }\bibfield  {title} {\enquote
  {\bibinfo {title} {Topological valley transport at bilayer graphene domain
  walls},}\ }\href {https://doi.org/10.1038/nature14364} {\bibfield  {journal}
  {\bibinfo  {journal} {Nature}\ }\textbf {\bibinfo {volume} {520}},\ \bibinfo
  {pages} {650--655} (\bibinfo {year} {2015})}\BibitemShut {NoStop}%
\bibitem [{\citenamefont {Killi}\ \emph {et~al.}(2010)\citenamefont {Killi},
  \citenamefont {Wei}, \citenamefont {Affleck},\ and\ \citenamefont
  {Paramekanti}}]{Kil10}%
  \BibitemOpen
  \bibfield  {author} {\bibinfo {author} {\bibfnamefont {Matthew}\ \bibnamefont
  {Killi}}, \bibinfo {author} {\bibfnamefont {Tzu-Chieh}\ \bibnamefont {Wei}},
  \bibinfo {author} {\bibfnamefont {Ian}\ \bibnamefont {Affleck}}, \ and\
  \bibinfo {author} {\bibfnamefont {Arun}\ \bibnamefont {Paramekanti}},\
  }\bibfield  {title} {\enquote {\bibinfo {title} {Tunable luttinger liquid
  physics in biased bilayer graphene},}\ }\href {\doibase
  10.1103/PhysRevLett.104.216406} {\bibfield  {journal} {\bibinfo  {journal}
  {Phys. Rev. Lett.}\ }\textbf {\bibinfo {volume} {104}},\ \bibinfo {pages}
  {216406} (\bibinfo {year} {2010})}\BibitemShut {NoStop}%
\bibitem [{\citenamefont {Killi}\ \emph {et~al.}(2011)\citenamefont {Killi},
  \citenamefont {Wu},\ and\ \citenamefont {Paramekanti}}]{Kil11}%
  \BibitemOpen
  \bibfield  {author} {\bibinfo {author} {\bibfnamefont {Matthew}\ \bibnamefont
  {Killi}}, \bibinfo {author} {\bibfnamefont {Si}~\bibnamefont {Wu}}, \ and\
  \bibinfo {author} {\bibfnamefont {Arun}\ \bibnamefont {Paramekanti}},\
  }\bibfield  {title} {\enquote {\bibinfo {title} {Band structures of bilayer
  graphene superlattices},}\ }\href {\doibase 10.1103/PhysRevLett.107.086801}
  {\bibfield  {journal} {\bibinfo  {journal} {Phys. Rev. Lett.}\ }\textbf
  {\bibinfo {volume} {107}},\ \bibinfo {pages} {086801} (\bibinfo {year}
  {2011})}\BibitemShut {NoStop}%
\bibitem [{\citenamefont {Li}\ \emph {et~al.}(2016)\citenamefont {Li},
  \citenamefont {Wang}, \citenamefont {McFaul}, \citenamefont {Zern},
  \citenamefont {Ren}, \citenamefont {Watanabe}, \citenamefont {Taniguchi},
  \citenamefont {Qiao},\ and\ \citenamefont {Zhu}}]{Li16}%
  \BibitemOpen
  \bibfield  {author} {\bibinfo {author} {\bibfnamefont {Jing}\ \bibnamefont
  {Li}}, \bibinfo {author} {\bibfnamefont {Ke}~\bibnamefont {Wang}}, \bibinfo
  {author} {\bibfnamefont {Kenton~J.}\ \bibnamefont {McFaul}}, \bibinfo
  {author} {\bibfnamefont {Zachary}\ \bibnamefont {Zern}}, \bibinfo {author}
  {\bibfnamefont {Yafei}\ \bibnamefont {Ren}}, \bibinfo {author} {\bibfnamefont
  {Kenji}\ \bibnamefont {Watanabe}}, \bibinfo {author} {\bibfnamefont
  {Takashi}\ \bibnamefont {Taniguchi}}, \bibinfo {author} {\bibfnamefont
  {Zhenhua}\ \bibnamefont {Qiao}}, \ and\ \bibinfo {author} {\bibfnamefont
  {Jun}\ \bibnamefont {Zhu}},\ }\bibfield  {title} {\enquote {\bibinfo {title}
  {Gate-controlled topological conducting channels in bilayer graphene},}\
  }\href {\doibase 10.1038/nnano.2016.158} {\bibfield  {journal} {\bibinfo
  {journal} {Nature Nanotechnology}\ }\textbf {\bibinfo {volume} {11}},\
  \bibinfo {pages} {1060--1065} (\bibinfo {year} {2016})}\BibitemShut {NoStop}%
\bibitem [{\citenamefont {Chen}\ \emph {et~al.}(2020)\citenamefont {Chen},
  \citenamefont {Zhou}, \citenamefont {Liu}, \citenamefont {Qiao},
  \citenamefont {Oezyilmaz},\ and\ \citenamefont {Martin}}]{Chen20}%
  \BibitemOpen
  \bibfield  {author} {\bibinfo {author} {\bibfnamefont {H.}~\bibnamefont
  {Chen}}, \bibinfo {author} {\bibfnamefont {P.}~\bibnamefont {Zhou}}, \bibinfo
  {author} {\bibfnamefont {J.}~\bibnamefont {Liu}}, \bibinfo {author}
  {\bibfnamefont {J.}~\bibnamefont {Qiao}}, \bibinfo {author} {\bibfnamefont
  {B.}~\bibnamefont {Oezyilmaz}}, \ and\ \bibinfo {author} {\bibfnamefont
  {J.}~\bibnamefont {Martin}},\ }\bibfield  {title} {\enquote {\bibinfo {title}
  {Gate controlled valley polarizer in bilayer graphene},}\ }\href {\doibase
  10.1038/s41467-020-15117-y} {\bibfield  {journal} {\bibinfo  {journal}
  {Nature Communications}\ }\textbf {\bibinfo {volume} {11}},\ \bibinfo {pages}
  {1202} (\bibinfo {year} {2020})}\BibitemShut {NoStop}%
\bibitem [{\citenamefont {Xiao}\ \emph {et~al.}(2007)\citenamefont {Xiao},
  \citenamefont {Yao},\ and\ \citenamefont {Niu}}]{Xia07}%
  \BibitemOpen
  \bibfield  {author} {\bibinfo {author} {\bibfnamefont {Di}~\bibnamefont
  {Xiao}}, \bibinfo {author} {\bibfnamefont {Wang}\ \bibnamefont {Yao}}, \ and\
  \bibinfo {author} {\bibfnamefont {Qian}\ \bibnamefont {Niu}},\ }\bibfield
  {title} {\enquote {\bibinfo {title} {Valley-contrasting physics in graphene:
  Magnetic moment and topological transport},}\ }\href {\doibase
  10.1103/PhysRevLett.99.236809} {\bibfield  {journal} {\bibinfo  {journal}
  {Phys. Rev. Lett.}\ }\textbf {\bibinfo {volume} {99}},\ \bibinfo {pages}
  {236809} (\bibinfo {year} {2007})}\BibitemShut {NoStop}%
\bibitem [{\citenamefont {Qiao}\ \emph {et~al.}(2011)\citenamefont {Qiao},
  \citenamefont {Jung}, \citenamefont {Niu},\ and\ \citenamefont
  {MacDonald}}]{Qiao11}%
  \BibitemOpen
  \bibfield  {author} {\bibinfo {author} {\bibfnamefont {Zhenhua}\ \bibnamefont
  {Qiao}}, \bibinfo {author} {\bibfnamefont {Jeil}\ \bibnamefont {Jung}},
  \bibinfo {author} {\bibfnamefont {Qian}\ \bibnamefont {Niu}}, \ and\ \bibinfo
  {author} {\bibfnamefont {Allan~H.}\ \bibnamefont {MacDonald}},\ }\bibfield
  {title} {\enquote {\bibinfo {title} {Electronic highways in bilayer
  graphene},}\ }\href {\doibase 10.1021/nl201941f} {\bibfield  {journal}
  {\bibinfo  {journal} {Nano Letters}\ }\textbf {\bibinfo {volume} {11}},\
  \bibinfo {pages} {3453--3459} (\bibinfo {year} {2011})}\BibitemShut {NoStop}%
\bibitem [{\citenamefont {Li}\ \emph {et~al.}(2018)\citenamefont {Li},
  \citenamefont {Zhang}, \citenamefont {Yin}, \citenamefont {Zhang},
  \citenamefont {Watanabe}, \citenamefont {Taniguchi}, \citenamefont {Liu},\
  and\ \citenamefont {Zhu}}]{Li18}%
  \BibitemOpen
  \bibfield  {author} {\bibinfo {author} {\bibfnamefont {Jing}\ \bibnamefont
  {Li}}, \bibinfo {author} {\bibfnamefont {Rui-Xing}\ \bibnamefont {Zhang}},
  \bibinfo {author} {\bibfnamefont {Zhenxi}\ \bibnamefont {Yin}}, \bibinfo
  {author} {\bibfnamefont {Jianxiao}\ \bibnamefont {Zhang}}, \bibinfo {author}
  {\bibfnamefont {Kenji}\ \bibnamefont {Watanabe}}, \bibinfo {author}
  {\bibfnamefont {Takashi}\ \bibnamefont {Taniguchi}}, \bibinfo {author}
  {\bibfnamefont {Chaoxing}\ \bibnamefont {Liu}}, \ and\ \bibinfo {author}
  {\bibfnamefont {Jun}\ \bibnamefont {Zhu}},\ }\bibfield  {title} {\enquote
  {\bibinfo {title} {A valley valve and electron beam splitter},}\ }\href
  {\doibase 10.1126/science.aao5989} {\bibfield  {journal} {\bibinfo  {journal}
  {Science}\ }\textbf {\bibinfo {volume} {362}},\ \bibinfo {pages} {1149--1152}
  (\bibinfo {year} {2018})}\BibitemShut {NoStop}%
\bibitem [{\citenamefont {Cheng}\ \emph {et~al.}(2018)\citenamefont {Cheng},
  \citenamefont {Liu}, \citenamefont {Jiang}, \citenamefont {Sun},\ and\
  \citenamefont {Xie}}]{Cheng18}%
  \BibitemOpen
  \bibfield  {author} {\bibinfo {author} {\bibfnamefont {Shu-guang}\
  \bibnamefont {Cheng}}, \bibinfo {author} {\bibfnamefont {Haiwen}\
  \bibnamefont {Liu}}, \bibinfo {author} {\bibfnamefont {Hua}\ \bibnamefont
  {Jiang}}, \bibinfo {author} {\bibfnamefont {Qing-Feng}\ \bibnamefont {Sun}},
  \ and\ \bibinfo {author} {\bibfnamefont {X.~C.}\ \bibnamefont {Xie}},\
  }\bibfield  {title} {\enquote {\bibinfo {title} {Manipulation and
  characterization of the valley-polarized topological kink states in
  graphene-based interferometers},}\ }\href {\doibase
  10.1103/PhysRevLett.121.156801} {\bibfield  {journal} {\bibinfo  {journal}
  {Phys. Rev. Lett.}\ }\textbf {\bibinfo {volume} {121}},\ \bibinfo {pages}
  {156801} (\bibinfo {year} {2018})}\BibitemShut {NoStop}%
\bibitem [{\citenamefont {Xavier}\ \emph {et~al.}(2010)\citenamefont {Xavier},
  \citenamefont {Pereira}, \citenamefont {Chaves}, \citenamefont {Farias},\
  and\ \citenamefont {Peeters}}]{xavier10}%
  \BibitemOpen
  \bibfield  {author} {\bibinfo {author} {\bibfnamefont {L.~J.~P.}\
  \bibnamefont {Xavier}}, \bibinfo {author} {\bibfnamefont {J.~M.}\
  \bibnamefont {Pereira}}, \bibinfo {author} {\bibfnamefont {Andrey}\
  \bibnamefont {Chaves}}, \bibinfo {author} {\bibfnamefont {G.~A.}\
  \bibnamefont {Farias}}, \ and\ \bibinfo {author} {\bibfnamefont {F.~M.}\
  \bibnamefont {Peeters}},\ }\bibfield  {title} {\enquote {\bibinfo {title}
  {Topological confinement in graphene bilayer quantum rings},}\ }\href
  {\doibase 10.1063/1.3431618} {\bibfield  {journal} {\bibinfo  {journal}
  {Applied Physics Letters}\ }\textbf {\bibinfo {volume} {96}},\ \bibinfo
  {pages} {212108} (\bibinfo {year} {2010})}\BibitemShut {NoStop}%
\bibitem [{\citenamefont {{da Costa}}\ \emph {et~al.}(2014)\citenamefont {{da
  Costa}}, \citenamefont {Zarenia}, \citenamefont {Chaves}, \citenamefont
  {Farias},\ and\ \citenamefont {Peeters}}]{Dacosta14}%
  \BibitemOpen
  \bibfield  {author} {\bibinfo {author} {\bibfnamefont {D.R.}\ \bibnamefont
  {{da Costa}}}, \bibinfo {author} {\bibfnamefont {M.}~\bibnamefont {Zarenia}},
  \bibinfo {author} {\bibfnamefont {Andrey}\ \bibnamefont {Chaves}}, \bibinfo
  {author} {\bibfnamefont {G.A.}\ \bibnamefont {Farias}}, \ and\ \bibinfo
  {author} {\bibfnamefont {F.M.}\ \bibnamefont {Peeters}},\ }\bibfield  {title}
  {\enquote {\bibinfo {title} {Analytical study of the energy levels in bilayer
  graphene quantum dots},}\ }\href {\doibase
  https://doi.org/10.1016/j.carbon.2014.06.078} {\bibfield  {journal} {\bibinfo
   {journal} {Carbon}\ }\textbf {\bibinfo {volume} {78}},\ \bibinfo {pages}
  {392--400} (\bibinfo {year} {2014})}\BibitemShut {NoStop}%
\bibitem [{\citenamefont {Serra}(2013)}]{Serra13}%
  \BibitemOpen
  \bibfield  {author} {\bibinfo {author} {\bibfnamefont {Lloren{\c{c}}}\
  \bibnamefont {Serra}},\ }\bibfield  {title} {\enquote {\bibinfo {title}
  {{M}ajorana modes and complex band structure of quantum wires},}\ }\href
  {\doibase 10.1103/PhysRevB.87.075440} {\bibfield  {journal} {\bibinfo
  {journal} {Phys. Rev. B}\ }\textbf {\bibinfo {volume} {87}},\ \bibinfo
  {pages} {075440} (\bibinfo {year} {2013})}\BibitemShut {NoStop}%
\bibitem [{\citenamefont {Osca}\ and\ \citenamefont {Serra}(2019)}]{Osca19}%
  \BibitemOpen
  \bibfield  {author} {\bibinfo {author} {\bibfnamefont {Javier}\ \bibnamefont
  {Osca}}\ and\ \bibinfo {author} {\bibfnamefont {Lloren{\c{c}}}\ \bibnamefont
  {Serra}},\ }\bibfield  {title} {\enquote {\bibinfo {title} {Complex
  band-structure analysis and topological physics of {M}ajorana nanowires},}\
  }\href {\doibase 10.1140/epjb/e2019-100011-2} {\bibfield  {journal} {\bibinfo
   {journal} {Eur. Phys. J. B}\ }\textbf {\bibinfo {volume} {92}},\ \bibinfo
  {pages} {101} (\bibinfo {year} {2019})}\BibitemShut {NoStop}%
\bibitem [{\citenamefont {Lehoucq}\ \emph {et~al.}(1998)\citenamefont
  {Lehoucq}, \citenamefont {Sorensen},\ and\ \citenamefont {Yang}}]{arpack}%
  \BibitemOpen
  \bibfield  {author} {\bibinfo {author} {\bibfnamefont {R.~B.}\ \bibnamefont
  {Lehoucq}}, \bibinfo {author} {\bibfnamefont {D.~C.}\ \bibnamefont
  {Sorensen}}, \ and\ \bibinfo {author} {\bibfnamefont {C.}~\bibnamefont
  {Yang}},\ }\href@noop {} {\emph {\bibinfo {title} {ARPACK Users Guide:
  Solution of Large-Scale Eigenvalue Problems with Implicitly Restarted Arnoldi
  Methods}}}\ (\bibinfo  {publisher} {Philadelphia: SIAM. ISBN
  978-0-89871-407-4},\ \bibinfo {year} {1998})\BibitemShut {NoStop}%
\bibitem [{\citenamefont {Susskind}(1977)}]{Susskind77}%
  \BibitemOpen
  \bibfield  {author} {\bibinfo {author} {\bibfnamefont {Leonard}\ \bibnamefont
  {Susskind}},\ }\bibfield  {title} {\enquote {\bibinfo {title} {Lattice
  fermions},}\ }\href {\doibase 10.1103/PhysRevD.16.3031} {\bibfield  {journal}
  {\bibinfo  {journal} {Phys. Rev. D}\ }\textbf {\bibinfo {volume} {16}},\
  \bibinfo {pages} {3031--3039} (\bibinfo {year} {1977})}\BibitemShut {NoStop}%
\bibitem [{\citenamefont {Nielsen}\ and\ \citenamefont
  {Ninomiya}(1981)}]{Nielsen81}%
  \BibitemOpen
  \bibfield  {author} {\bibinfo {author} {\bibfnamefont {H.B.}\ \bibnamefont
  {Nielsen}}\ and\ \bibinfo {author} {\bibfnamefont {M.}~\bibnamefont
  {Ninomiya}},\ }\bibfield  {title} {\enquote {\bibinfo {title} {Absence of
  neutrinos on a lattice: (i). proof by homotopy theory},}\ }\href {\doibase
  https://doi.org/10.1016/0550-3213(81)90361-8} {\bibfield  {journal} {\bibinfo
   {journal} {Nuclear Physics B}\ }\textbf {\bibinfo {volume} {185}},\ \bibinfo
  {pages} {20--40} (\bibinfo {year} {1981})}\BibitemShut {NoStop}%
\bibitem [{\citenamefont {Hern\'andez}\ and\ \citenamefont
  {Lewenkopf}(2012)}]{Lewe12}%
  \BibitemOpen
  \bibfield  {author} {\bibinfo {author} {\bibfnamefont {Alexis~R.}\
  \bibnamefont {Hern\'andez}}\ and\ \bibinfo {author} {\bibfnamefont {Caio~H.}\
  \bibnamefont {Lewenkopf}},\ }\bibfield  {title} {\enquote {\bibinfo {title}
  {Finite-difference method for transport of two-dimensional massless {D}irac
  fermions in a ribbon geometry},}\ }\href {\doibase
  10.1103/PhysRevB.86.155439} {\bibfield  {journal} {\bibinfo  {journal} {Phys.
  Rev. B}\ }\textbf {\bibinfo {volume} {86}},\ \bibinfo {pages} {155439}
  (\bibinfo {year} {2012})}\BibitemShut {NoStop}%
\bibitem [{\citenamefont {Park}(2019)}]{Par19}%
  \BibitemOpen
  \bibfield  {author} {\bibinfo {author} {\bibfnamefont {Changsoo}\
  \bibnamefont {Park}},\ }\bibfield  {title} {\enquote {\bibinfo {title}
  {Magnetoelectrically controlled valley filter and valley valve in bilayer
  graphene},}\ }\href {\doibase 10.1103/PhysRevApplied.11.044033} {\bibfield
  {journal} {\bibinfo  {journal} {Phys. Rev. Applied}\ }\textbf {\bibinfo
  {volume} {11}},\ \bibinfo {pages} {044033} (\bibinfo {year}
  {2019})}\BibitemShut {NoStop}%
\bibitem [{\citenamefont {Recher}\ \emph {et~al.}(2009)\citenamefont {Recher},
  \citenamefont {Nilsson}, \citenamefont {Burkard},\ and\ \citenamefont
  {Trauzettel}}]{Recher09}%
  \BibitemOpen
  \bibfield  {author} {\bibinfo {author} {\bibfnamefont {Patrik}\ \bibnamefont
  {Recher}}, \bibinfo {author} {\bibfnamefont {Johan}\ \bibnamefont {Nilsson}},
  \bibinfo {author} {\bibfnamefont {Guido}\ \bibnamefont {Burkard}}, \ and\
  \bibinfo {author} {\bibfnamefont {Bj\"orn}\ \bibnamefont {Trauzettel}},\
  }\bibfield  {title} {\enquote {\bibinfo {title} {Bound states and magnetic
  field induced valley splitting in gate-tunable graphene quantum dots},}\
  }\href {\doibase 10.1103/PhysRevB.79.085407} {\bibfield  {journal} {\bibinfo
  {journal} {Phys. Rev. B}\ }\textbf {\bibinfo {volume} {79}},\ \bibinfo
  {pages} {085407} (\bibinfo {year} {2009})}\BibitemShut {NoStop}%
\bibitem [{\citenamefont {Li}\ \emph {et~al.}(2011)\citenamefont {Li},
  \citenamefont {Martin}, \citenamefont {B\"uttiker},\ and\ \citenamefont
  {Morpurgo}}]{Li11}%
  \BibitemOpen
  \bibfield  {author} {\bibinfo {author} {\bibfnamefont {Jian}\ \bibnamefont
  {Li}}, \bibinfo {author} {\bibfnamefont {Ivan}\ \bibnamefont {Martin}},
  \bibinfo {author} {\bibfnamefont {Markus}\ \bibnamefont {B\"uttiker}}, \ and\
  \bibinfo {author} {\bibfnamefont {Alberto~F.}\ \bibnamefont {Morpurgo}},\
  }\bibfield  {title} {\enquote {\bibinfo {title} {Topological origin of subgap
  conductance in insulating bilayer graphene},}\ }\href@noop {} {\bibfield
  {journal} {\bibinfo  {journal} {Nature Phys.}\ }\textbf {\bibinfo {volume}
  {7}},\ \bibinfo {pages} {38} (\bibinfo {year} {2011})}\BibitemShut {NoStop}%
\bibitem [{\citenamefont {Pan}\ \emph {et~al.}(2014)\citenamefont {Pan},
  \citenamefont {Li}, \citenamefont {Liu}, \citenamefont {Zhu}, \citenamefont
  {Qiao},\ and\ \citenamefont {Yao}}]{Pan14}%
  \BibitemOpen
  \bibfield  {author} {\bibinfo {author} {\bibfnamefont {Hui}\ \bibnamefont
  {Pan}}, \bibinfo {author} {\bibfnamefont {Zhenshan}\ \bibnamefont {Li}},
  \bibinfo {author} {\bibfnamefont {Cheng-Cheng}\ \bibnamefont {Liu}}, \bibinfo
  {author} {\bibfnamefont {Guobao}\ \bibnamefont {Zhu}}, \bibinfo {author}
  {\bibfnamefont {Zhenhua}\ \bibnamefont {Qiao}}, \ and\ \bibinfo {author}
  {\bibfnamefont {Yugui}\ \bibnamefont {Yao}},\ }\bibfield  {title} {\enquote
  {\bibinfo {title} {Valley-polarized quantum anomalous hall effect in
  silicene},}\ }\href {\doibase 10.1103/PhysRevLett.112.106802} {\bibfield
  {journal} {\bibinfo  {journal} {Phys. Rev. Lett.}\ }\textbf {\bibinfo
  {volume} {112}},\ \bibinfo {pages} {106802} (\bibinfo {year}
  {2014})}\BibitemShut {NoStop}%
\bibitem [{\citenamefont {Lu}\ \emph {et~al.}(2017)\citenamefont {Lu},
  \citenamefont {Qiu}, \citenamefont {Ye}, \citenamefont {Fan}, \citenamefont
  {Ke}, \citenamefont {Zhang},\ and\ \citenamefont {Liu}}]{Lu17}%
  \BibitemOpen
  \bibfield  {author} {\bibinfo {author} {\bibfnamefont {Jiuyang}\ \bibnamefont
  {Lu}}, \bibinfo {author} {\bibfnamefont {Chunyin}\ \bibnamefont {Qiu}},
  \bibinfo {author} {\bibfnamefont {Liping}\ \bibnamefont {Ye}}, \bibinfo
  {author} {\bibfnamefont {Xiying}\ \bibnamefont {Fan}}, \bibinfo {author}
  {\bibfnamefont {Manzhu}\ \bibnamefont {Ke}}, \bibinfo {author} {\bibfnamefont
  {Fan}\ \bibnamefont {Zhang}}, \ and\ \bibinfo {author} {\bibfnamefont
  {Zhengyou}\ \bibnamefont {Liu}},\ }\bibfield  {title} {\enquote {\bibinfo
  {title} {Observation of topological valley transport of sound in sonic
  crystals},}\ }\href@noop {} {\bibfield  {journal} {\bibinfo  {journal}
  {Nature Physics}\ }\textbf {\bibinfo {volume} {13}},\ \bibinfo {pages} {369}
  (\bibinfo {year} {2017})}\BibitemShut {NoStop}%
\bibitem [{\citenamefont {Gao}\ \emph {et~al.}(2017)\citenamefont {Gao},
  \citenamefont {Xue}, \citenamefont {Yang}, \citenamefont {Lai}, \citenamefont
  {Yu}, \citenamefont {Lin}, \citenamefont {Chong},\ and\ \citenamefont
  {Shvets}}]{Gao17}%
  \BibitemOpen
  \bibfield  {author} {\bibinfo {author} {\bibfnamefont {Fei}\ \bibnamefont
  {Gao}}, \bibinfo {author} {\bibfnamefont {Haoran}\ \bibnamefont {Xue}},
  \bibinfo {author} {\bibfnamefont {Zhaoju}\ \bibnamefont {Yang}}, \bibinfo
  {author} {\bibfnamefont {Kueifu}\ \bibnamefont {Lai}}, \bibinfo {author}
  {\bibfnamefont {Yang}\ \bibnamefont {Yu}}, \bibinfo {author} {\bibfnamefont
  {Xiao}\ \bibnamefont {Lin}}, \bibinfo {author} {\bibfnamefont {Yidong}\
  \bibnamefont {Chong}}, \ and\ \bibinfo {author} {\bibfnamefont {Gennady}\
  \bibnamefont {Shvets}},\ }\bibfield  {title} {\enquote {\bibinfo {title}
  {Topologically protected refraction of robust kink states in valley photonic
  crystals},}\ }\href@noop {} {\bibfield  {journal} {\bibinfo  {journal}
  {Nature Physics}\ }\textbf {\bibinfo {volume} {14}},\ \bibinfo {pages} {140}
  (\bibinfo {year} {2017})}\BibitemShut {NoStop}%
\end{thebibliography}%

\end{document}